%% file: sample631.tex
\documentclass{aastex631}

\usepackage{multirow}

\usepackage{makecell}
\usepackage{amsmath}

\shorttitle{Disconnecting the Dots}
\shortauthors{Zucker et al.}
\graphicspath{{./}{figures/}}

\begin{document}

\title{Disconnecting the Dots: Re-Examining the Nature of Stellar ``Strings" in the Milky Way}

\correspondingauthor{Catherine Zucker}
\email{czucker@stsci.edu}

\author[0000-0002-2250-730X]{Catherine Zucker}
\affiliation{Space Telescope Science Institute, 3700 San Martin Drive, Baltimore, MD 21218, USA}
\altaffiliation{Hubble Fellow}

\author[0000-0003-4797-7030]{J. E. G. Peek}
\affiliation{Space Telescope Science Institute, 3700 San Martin Drive, Baltimore, MD 21218, USA}
\affiliation{Department of Physics \& Astronomy, Johns Hopkins University, Baltimore, MD 21218, USA}

\author[0000-0003-3217-5967]{Sarah Loebman}
\affiliation{Department of Physics, University of California, Merced, 5200 Lake Road, Merced, CA 95343, USA}



\begin{abstract}
Recent analyses of \textit{Gaia} data have resulted in the identification of new stellar structures, including a new class of extended stellar filaments called stellar ``strings", first proposed by \citet{Kounkel_2019}. We explore the spatial, kinematic, and chemical composition of strings to demonstrate that these newfound structures are largely inconsistent with being physical objects whose members share a common origin. Examining the 3D spatial distribution of string members, we find that the spatial dispersion around the claimed string spine does not improve in the latest \textit{Gaia} DR3 data release --- despite tangible gains in the signal-to-noise of the parallax measurements --- counter to expectations of a bona fide structure. Using the radial velocity dispersion of the strings (averaging $\rm\sigma_{V_r} = 16\;km\;s^{-1}$) to estimate their virial masses, we find that all strings are gravitationally unbound. Given the finding that the strings are dispersing, the reported stellar ages of the strings are typically $120\times$ larger than their measured dispersal times. Finally, we validate prior work that stellar strings are more chemically homogeneous than their local field stars, but show it is possible to obtain the same signatures of chemical homogeneity by drawing random samples of stars from spatially, temporally, and kinematically unrelated open clusters. Our results show that while some strings may be composed of real sub-structures, there is no consistent evidence for larger string-like connections over the sample. These results underline the need for caution in over-interpreting the significance of these strings and their role in understanding the star formation history of the Milky Way.

\end{abstract}

\keywords{}


\section{Introduction} \label{sec:intro}
It is commonly accepted that most stars in the Milky Way were born in close proximity to other stars, constituting a stellar structure that formed at the same time within the same parental molecular gas structure \citep[e.g.][]{Lada_2003}. These stellar siblings should be similar to one another in terms of their location, age, kinematics, and chemistry. As these stellar structures dissolve into the Galactic field, they offer an opportunity to study the star formation history of the Milky Way and the chemodynamical evolution of its disk. As the largest and most accurate astrometric catalog of stars ever produced, \textit{Gaia} \citep{Gaia_2016} offers an unprecedented opportunity to study these stellar structures from their formation to their dissolution. By constraining the distances and proper motions to over a billion stars, as well as the radial velocities of millions of stars, \textit{Gaia} has not only shed new light on the spatial and dynamical properties of existing stellar structures, but has also enabled the discovery of new ones. These discoveries include hundreds of previously unknown open clusters \citep[e.g.][]{Castro_Ginard_2020}, as well as new classes of stellar structures with much more extended spatial distributions \citep[see e.g. discussion in \S 3.5 of][]{Cantat_Gaudin_2022}, including stellar ``streams" in the Galactic disk \citep{Meingast_2019}, extended stellar coronae \citep{Meingast_2021, Ratzenbock_2020, Meingast_2019_Hyades}, stellar ``pearls" \citep{Coronado_2022}, stellar ``relic filaments" \citep{Jerabkova_2019,Beccari_2020}, stellar ``snakes" \citep{Wang_2022}, and stellar ``strings" \citep[][]{Kounkel_2019} \citep[see also][]{Kounkel_2020}.

In Table \ref{tab:litcomp}, we compare and contrast the properties of these proposed extended stellar structures, including their velocity dispersions, claimed co-evality, gravitational boundedness, lengths, and widths. All extended structures have similar lengths (spanning $\approx$ 100 - 400 pc) and widths ($\approx$ few tens of parsecs) with typical aspect ratios between 3:1 and 10:1. None of the studies in Table \ref{tab:litcomp} explicitly require the structures to be gravitationally bound, though \citet{Kounkel_2019} argue that strings are ``weakly bound" while \citet{Meingast_2019} and \citet{Meingast_2021} argue that the recently discovered streams in the disk and extended coronae around open clusters are gravitationally unbound. All structures are argued to be co-eval with the exception of stellar ``pearls", which are distinct clusters following similar orbits in the Galaxy that manifest as overdensities in action-angle space \citep{Coronado_2022}.  Stellar strings are similar to stellar ``snakes" and stellar ``relic filaments" in that all three structures are identified in part via clustering algorithms in 5D space (e.g. HDBSCAN, DBSCAN, Friends-of-Friends) and have radial velocity dispersions spanning $\rm \approx 5 \; km\; s^{-1}$ (relic filaments) to $\rm \approx 15 \; km\; s^{-1}$ (strings). The filamentary structure seen in strings, snakes, and relic filaments is argued to be primordial, likely forming in elongated giant molecular filaments \citep{Goodman_2014, Zucker_2015, Zucker_2018, Ragan_2014, Wang_2015, Zucker_2019}. In contrast, stellar streams in both the disk \citep{Meingast_2019} and the halo \citep{Gialluca_2021,Kuzma_2015}, as well as recently discovered extended stellar coronae \citep{Meingast_2021}, have much smaller velocity dispersions ($\rm \approx 1 \; km \; s^{-1}$) with the filamentary structure forming as a result of dynamical tidal forces dissolving a central cluster. 

There are four attributes that members of a newly discovered stellar structure (like those in Table \ref{tab:litcomp}) should share in order to plausibly be considered co-eval, or born at the same time within the same parental molecular gas structure. First, stars in a structure should have largely similar ages. Second, members of a stellar structure should be close enough to one another in 3D space such that they could have been born in the same location. Third, members of a stellar structure should share similar motions, as evidenced by small dispersion in their \textit{Gaia} tangential and radial velocities. Finally, members of a stellar structure should have similar metallicities, as evidenced by small dispersion in elemental abundances \citep[e.g. as measured by spectroscopic surveys like GALAH and APOGEE;][]{GALAH_DR3, APOGEE_DR16}. 

With these attributes in mind, we take a closer look at the spatial, kinematic, and abundance variations of the most extreme population in Table 1 -- the stellar strings -- first proposed by \citet{Kounkel_2019} (hereafter KC19). KC19 present a sample of 328 claimed co-eval stellar strings. While KC19 do not quantitatively define a string, they argue that strings should appear to be filamentary, roughly parallel to the Galactic plane, coherent both spatially and kinematically, and that their stellar members can generally be characterized with a single isochrone. The median projected length and width of a string is 190 pc and 30 pc, respectively. 

KC19 identify the strings in a multi-step process. First, they apply the HDBSCAN algorithm \citep{HDBSCAN} in 5D space ($l$, $b$, parallax $\pi$, and proper motions) to a sample of stars out to 1 kpc from the Sun detected in \textit{Gaia} DR2. Specifically, they perform several iterations of the HDBSCAN algorithm over different parallax ranges, primarily with the ``leaf" clustering method, to obtain a set of stellar groups with similar 5D properties. Then the authors manually merge and split the groups detected in the various iterations by hand. Next, KC19 assign an age to each group using a combination of isochrone fitting and a convolutional neural network. Finally, once a sample of stellar groups is identified via HDBSCAN, KC19 manually assemble the strings by either i.) connecting the individual groups with similar ages using the tool TOPCAT \citep{topcat} or ii) deciding that a single group possesses enough filamentary morphology to be classified as a string. Finally, KC19 visually check that the strings are ``fully continuous [and] coherent in all kinematic [i.e. tangential velocities] and spatial [i.e. $l$, $b$, $\pi$] dimensions." After the groups are connected, KC19 compute a ``spine" for the string in 5D space by averaging the star-by-star ($l$, $b$, $\pi$, and kinematics) results in different plane-of-the-sky longitude bins along the projected string, before smoothing with a Savitzky-Golay filter to avoid strong fluctuations in the averages. 

In this work, we independently test the kinematic, spatial, and chemical coherence of the stellar strings using data not fully considered by and/or available at the time of KC19. In \S \ref{data} we present the publicly available spatial and kinematic data for stellar strings from \textit{Gaia} DR2 and DR3 utilized in this work, along with ancillary spectroscopic data used to examine the elemental abundance variations within a subset of the strings. In \S \ref{methods_results} we use these data to derive estimates of the stars' 3D spatial dispersion around their respective string ``spines", their radial velocity dispersions, their predicted virial masses, their predicted dynamical lifetimes, and their elemental abundance variations. We then use these constraints to show that nearly all of these stellar strings are inconsistent with being co-eval, physical entities, and are rather artificial structures affected by limitations in the manual assembly process used in their selection. In \S \ref{discussion} we discuss the implications of the strings' nonphysical nature within the wider context of the \textit{Gaia} literature on extended stellar structures. Finally, we conclude in \S \ref{conclusions}.

\begin{deluxetable*}{ccccccccc}
\tablecaption{Proposed Extended Stellar Structures in the Galactic Disk\label{tab:litcomp}}
\tabletypesize{\scriptsize}
\setlength{\tabcolsep}{2pt}
\def\arraystretch{1.75}
\colnumbers
\tablehead{
\colhead{Name}  & \colhead{ID} & \colhead{Claimed Co-Eval?} & \colhead{$\sigma_{V_r}$} & \colhead{Length} & \colhead{Width} & \colhead{Bound?} & \colhead{Formation} & \colhead{Publication}\\
\colhead{} & \colhead{} & \colhead{} & \colhead{} & \colhead{} & \colhead{} & \colhead{} & \colhead{} & \colhead{}
}
\startdata
String	&	HDBSCAN (5D)	&	Yes	& 15	& 200	&	30	&	Weakly Bound	&	Primordial Filaments & \makecell{\text{\citet{Kounkel_2019}} \\ \text{\citet{Kounkel_2020}}} 	\\
Snake	&	FoF (5D)	&	Yes	&	10 	& 200	&	90	&	Not claimed	&	Primordial Filaments & \makecell{\text{\citet{Tian_2020}} \\ \text{\citet{Wang_2022}}} 	\\
Relic Filament	&	DBSCAN (5D)	&	Yes	&	5 	&	90 - 260 &	10 - 50	&	Not claimed	&	Primordial Filaments &  \makecell{\text{\citet{Jerabkova_2019}} \\ \text{\citet{Beccari_2020}}} \\
Extended Corona		&	Convergent Point (Deproj. 5D) &	Yes	&	1 	& 100 - 400	& 50 - 75	&	 Unbound	&	Tidally Disrupted Clusters & \text{\citet{Meingast_2021}}*	\\
Stream	& Wavelet Decomposition (6D)	&	Yes	&	 1	&	400	&	50	&	Unbound	&	Tidally Disrupted Cluster & \makecell{\text{\citet{Meingast_2019}} \\ \text{\citet{Ratzenbock_2020}}} 	\\
Pearl	&	Orbit-Phase Space FoF (6D)	&	No	&	--	& --		&	--	&	Unbound	&	Primordial Filaments & \text{\citet{Coronado_2022}}	\\
\enddata
\tablecomments{Comparison of the identification, properties, and physical interpretation of extended stellar structures recently discovered in the \textit{Gaia} era. (1) Name of the extended stellar structure (2) \textit{Primary} identification method of the structure, specifying whether initial identification was made in 5D (not including radial velocities) or 6D (including radial velocities) space. See the references below for full details on the multi-step structure identification. (3) Whether the structures are claimed to be be co-eval in their original publications.  (4) Typical radial velocity dispersion of the structure. (5) Average length or the range of lengths over the sample. (6) Average width or range of widths over the sample. (7) Whether the structures are suggested to be bound or unbound in their original publications. (8) The proposed formation mechanism for the structures, specifically whether their filamentary morphology is argued to be primordial (in filamentary giant molecular clouds) or whether it is the result of tidal stretching dissolving a central cluster. (9) Original publications describing identification and properties of the proposed extended stellar structures. No length, width, or radial velocity dispersion are provided for the pearls, as they are not argued to be monolithic structures, but rather sets of distinct clusters that follow similar orbits.}
\tablenotetext{*}{See also \citet{Moranta_2022} for extended coronae identified in 5D space via HDBSCAN}
\end{deluxetable*}

\section{Data} \label{data}
KC19 identify 1312 stellar groups and 328 stellar strings. A group is a single set of stars identified in HDBSCAN with similar 5D properties ($l$, $b$, parallax $\pi$, and proper motions), while a string is either a collection of connected HDBSCAN groups or a single HDBSCAN group deemed to be filamentary in KC19. We only consider the stellar strings in this work. We obtain the \textit{Gaia} DR2 data \citep{Gaia_2018} (sky coordinates, parallaxes, and parallax errors) on the string stars directly from KC19 (see their Table 1), and we crossmatch their Table 1 with \textit{Gaia} DR3 \citep{Gaia_2021} to obtain updated constraints on the parallax and parallax errors of the string stars. The XYZ positions (the Heliocentric Galactic Cartesian Coordinates) of string spines (defined using \textit{Gaia} DR2 data) are obtained from Table 3 in KC19 and will be used to calculate the 3D dispersion of string members around their respective spines in \S \ref{subsec:spatial}. To analyze the kinematic coherence of the strings in \S \ref{subsec:dynamics}, we adopt the updated radial velocity measurements from \textit{Gaia} DR3, which provides radial velocity data for $\approx 5 \times$ more stars than available in \textit{Gaia} DR2 in KC19. To explore the metallicity distribution within the strings, we use the catalog from \citet[][hereafter MHM22]{Manea_2022}. MHM22 leverages GALAH DR3 \citep{GALAH_DR3}  elemental abundance measurements (e.g. \texttt{[Fe/H]}) and their reported uncertainties to analyze the chemical homogeneity of stars in nearby stellar structures, including ten strings (see their Supplementary Data). To compare the chemical homogeneity of the strings to a benchmark sample of open clusters, we adopt the catalog from \citet{Spina_2021} (see their Table 1), which compiles a similar set of elemental abundance measurements from GALAH \citep{GALAH_DR3} and APOGEE \citep{APOGEE_DR16} for a sample of stars in hundreds of open clusters across the Galactic disk.

\begin{figure*}
    \centering
    \includegraphics[width=1.0\textwidth]{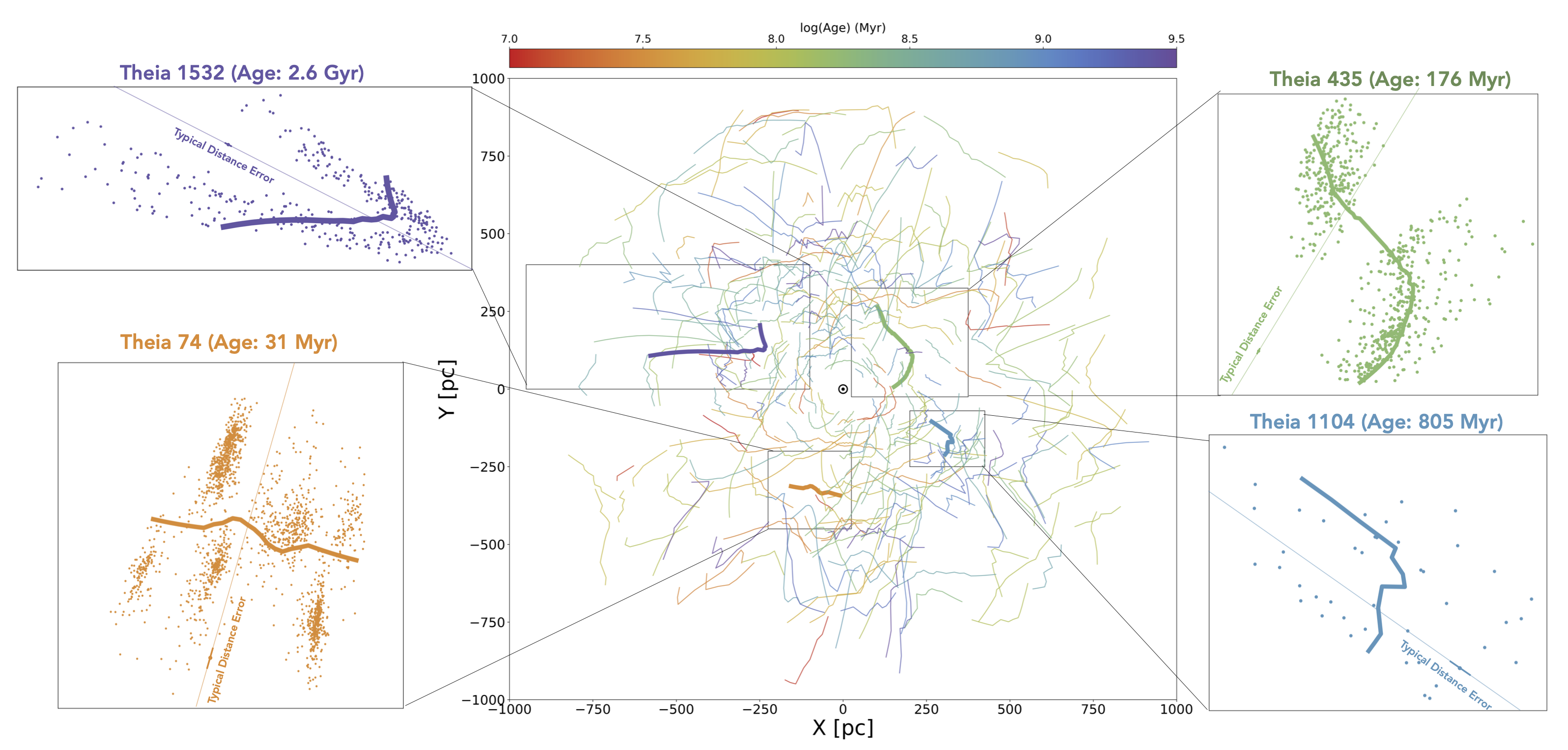}
    \caption{\textit{Center}: A topdown view of stellar strings, colored by age, based on KC19 (see their Figure 13). In the zoom-in boxes, we show the actual distribution of stellar members around the string ``spine" in \textit{Gaia} DR2 for a sub-sample of four strings spanning a range of ages. The typical distance error to stars is very small with N average parallax signal-to-noise $>$ 44 in \textit{Gaia} DR2. Since the strings ``spines" are derived using \textit{Gaia} DR2 data, these DR2 stellar distributions are the actual data used to derive the string properties in KC19. We find that these four strings are representative of the morphology seen across the full sample: the dispersion along the line of sight is significantly larger than the typical error on the parallax measurements (e.g. Theia 1532, Theia 1104), and many strings are composed of isolated groups with no evidence of connection between them (e.g. Theia 74, Theia 435). See Figure Set \ref{fig:figset} in the Appendix or the \href{https://faun.rc.fas.harvard.edu/czucker/Paper\_Figures/String_Gallery\_Interactive.html}{online interactive figure gallery} for similar panels for the remaining strings in the KC19 sample, as well as additional comparisons to the \textit{Gaia} DR3 stellar distributions. 
    \label{fig:topdown}}
\end{figure*}

\section{The Spatial, Dynamical, and Chemical Composition of Stellar Strings} \label{methods_results}

In this section, we re-examine the spatial (\S \ref{subsec:spatial}), dynamical (\S \ref{subsec:dynamics}), and chemical (\S \ref{subsec:chemical}) distribution of the strings in KC19 and present a summary of their derived properties in Table \ref{tab:summary}.\footnote{On \href{https://doi.org/10.5281/zenodo.6984474}{Zenodo}, we provide a Jupyter Notebook that reproduces all the results in this section, including the values in Table \ref{tab:summary}, and the data behind Figures \ref{fig:topdown}, \ref{fig:dispersion}, \ref{fig:age_vs_sigmaVR}, \ref{fig:lifetimes} and \ref{fig:metallicity} (doi:10.5281/zenodo.6984474).}

\subsection{3D Spatial Properties of Stellar Strings} \label{subsec:spatial}

In Figure \ref{fig:topdown} we show a topdown XY \textit{Gaia} DR2 view of the stellar strings as reproduced from KC19 (see their Figure 13). We highlight a selection of strings (over a range of ages) to convey the relationship between each string and its underlying stellar membership. Similar plots for the rest of the string sample are shown in Figure Set \ref{fig:figset} in the Appendix. The distances to the string stars are very well constrained: the median signal to noise of the parallax measurements per string surpasses 44:1 in DR2 and 61:1 in DR3. For a majority of the strings, the dispersion around the spine is much larger than the average distance uncertainty. Many strings are composed of discrete stellar groups that lack clear connections in 3D physical space, despite that interpretation in KC19.

\begin{figure*}
    \centering
    \includegraphics[width=1.0\textwidth]{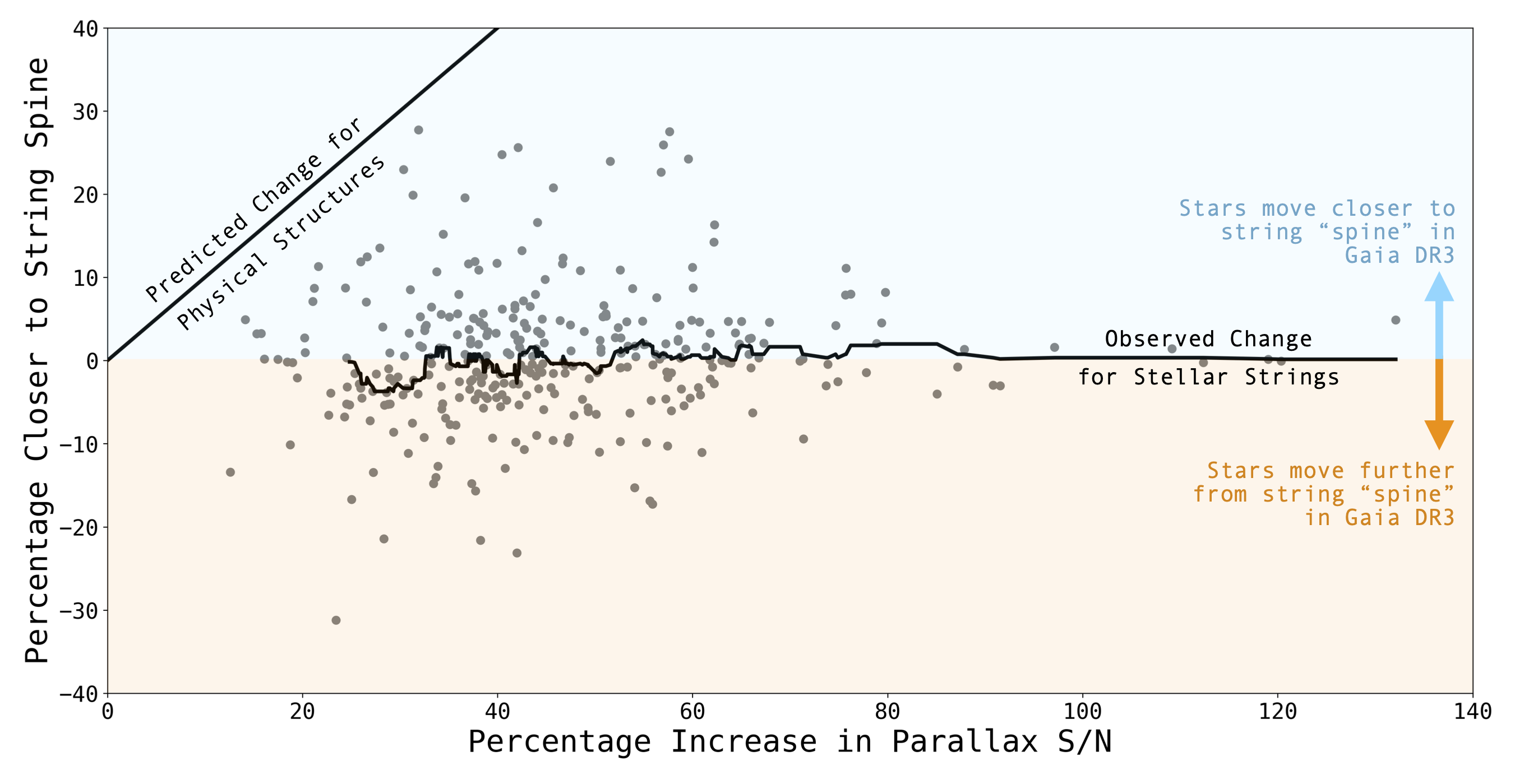}
    \caption{Change in the 3D spatial dispersion of stars around their string's spine from \textit{Gaia} DR2 to \textit{Gaia} DR3. Each grey dot represents one of the 328 individual strings. The dots show the average percentage change in the 3D spatial offset from the spine as a function of the average increase in the signal-to-noise of the parallax measurements for the stars. Positive percentage (blue region) indicates the stars move closer to their spine with improved parallax measurements, while negative percentage (orange region) indicates they move farther away. The black line shows the rolling median and indicates that despite a significant improvement in the signal-to-noise of the parallax measurements, the dispersion around the spine does not improve on average as expected of genuine spatial structures. The diagonal line in the top left shows the predicted trend for genuine space structures. 
    \label{fig:dispersion}}
\end{figure*}

Leveraging the improved astrometric precision of $Gaia$ DR3, we compare the 3D spatial dispersion of stars around the string spine in \textit{Gaia} DR2 versus \textit{Gaia} DR3 to determine whether the dispersion around the spine decreases as parallax errors improve, as expected of real structures. For each star, we compute its 3D offset from the string's spine in \textit{Gaia} DR2 and DR3, and then average the results per string. The results are presented in Figure \ref{fig:dispersion}, which shows the average percentage that the stars move closer to or further from the spine as a function of the increase in the signal to noise of the parallax measurements. Despite the signal to noise of the parallaxes improving by 20\%--120\% in \textit{Gaia} DR3, there is \textit{no improvement} in the stars' offsets from the spines. The lack of improvement in the stars' 3D spatial dispersion is inconsistent with the claim that strings are co-eval, physical entities whose members share a common origin. However, we do observe (c.f. Figure Set \ref{fig:figset} in the Appendix and the \href{https://faun.rc.fas.harvard.edu/czucker/Paper\_Figures/String\_Gallery\_Interactive.html}{online interactive figure gallery}) that in \textit{some} cases the 3D spatial dispersion within \textit{individual} stellar groups inside a string does improve. This suggests these strings may be partly composed of real stellar subgroups; however, we see no evidence for the larger string connections.

\subsection{The Dynamical Properties of Stellar Strings} \label{subsec:dynamics}
While the lack of improvement in the 3D spatial dispersion of stars around the string spines raises concerns about their fidelity, one way to validate the authenticity of the strings is to show that their stellar members still share similar motions. KC19 analyze the dispersion in the tangential velocities of string members and find them to be $\rm < 2.5 \; km \; s^{-1}$. We expect a small dispersion in the tangential velocities, since the stellar groups that were manually assembled into strings must share similar 5D properties ($l$, $b$, parallax $\pi$, and proper motions) to be detected with HDBSCAN. As such, the fairest way to evaluate the authenticity of the strings is to characterize their velocity dispersion in the sixth dimension --- the radial velocity dimension ---  not considered in the original 5D clustering algorithm. KC19 find that the dispersion in the radial velocities, $\rm \sigma_{V_r}$, span $\rm 5 - 40 \; km \; s^{-1}$ with an average radial velocity dispersion of $\rm 16 \; km \; s^{-1}$, a factor of $5-10\times$ larger than for the tangential velocities (see Figure 12 from KC19). The typical \textit{Gaia}-based radial velocity dispersion for loosely bound open clusters is $\rm \approx 1 \; km \; s^{-1}$ \citep{Soubiran_2018}, making these strings at least dynamically very atypical for known co-eval structures.

\begin{figure*}
    \centering
    \includegraphics[width=0.7\textwidth]{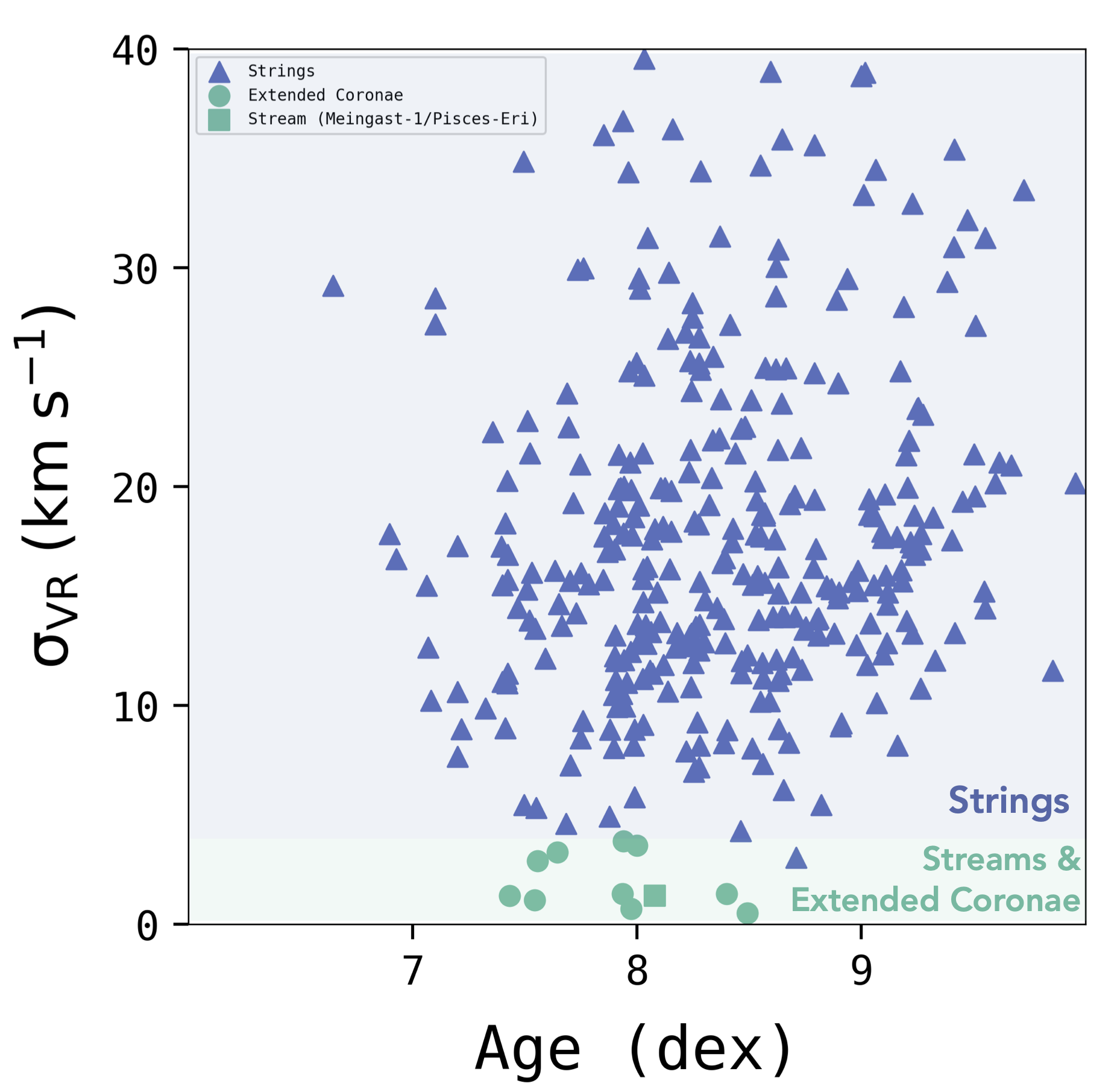}
    \caption{Error-weighted radial velocity dispersions of the strings as a function of their reported ages in KC19. Alongside the strings, we show comparable measurements for the extended stellar coronae \citep{Meingast_2021} around nearby open clusters (including e.g. the Pleaides and $\alpha$ Per) and the stellar streams \citep[Meingast-1/Pisces-Eridanus; see][]{Meingast_2019, Hawkins_2020} summarized in Table \ref{tab:litcomp}.
    \label{fig:age_vs_sigmaVR}}
\end{figure*}

Leveraging the expanded catalog of radial velocities in \textit{Gaia} DR3 --- providing $5\times$ more measurements than available at the time of KC19 --- we compute updated, error-weighted radial velocity dispersions for the strings. Before calculating the error-weighted radial velocity dispersion, we mask out stars classified as astrometric, spectroscopic, or eclipsing binaries in \textit{Gaia} DR3 by requiring the \texttt{non\_single\_star} flag to be 0. We adopt the same methodology used in \citet{Soubiran_2018} to calculate the error-weighted radial velocity dispersion for open clusters: 

\begin{equation} \label{eq:weighted_sigmaVR}
\sigma_{\rm V_{r}}^2=\frac{\underset{i}{\sum} w_i}{(\underset{i}{\sum} w_i)^2 - \underset{i}{\sum} w_i^2 }\,
\underset{i}{\sum} w_i ({\rm RV}_i -{\rm {RV_{str})}}^2
\end{equation}

where $w_i$ is the weight for the $i$th star ($w_i = \frac{1}{{RV_{err,i}}^{2}}$), $RV_{err,i}$ is the star's radial velocity error, $RV_i$ is the star's radial velocity measurement, and $RV_{str}$ is the error-weighted mean velocity of all stars in the string. Despite significantly enlarging the radial velocity sample, removing known binaries, and implementing error-weighting, the average string radial velocity dispersion stays the same at $\rm 16 \; km \; s^{-1}$. Figure \ref{fig:age_vs_sigmaVR} shows the distribution of error-weighted radial velocity dispersion for the strings as a function of their reported ages from KC19, alongside comparable measurements for extended stellar coronae and stellar streams in the disk (see Table \ref{tab:litcomp}).
 
Using the updated radial velocity dispersions, we estimate the predicted virial mass of each string as:  

\begin{equation}
M_{vir} = \frac{\sigma_{V_r}^{2} \times \eta \times r_{hm}}{G}
\end{equation}

where $r_{hm}$ is the adopted half-mass radius of the string. The parameter $\eta$ is a dimensionless constant that depends on the shape of the density profile, for which we very conservatively adopt $\eta = 1$.\footnote{The $\eta$ parameter is typically assumed to be $\approx 10$ for a Plummer model \citep{Plummer_1911} characterized by steep density profiles. Since recent studies have shown that $\eta$ can be smaller (consistent with much broader density profiles), particularly for younger systems due to e.g. mass segregation \citep{Zwart_2010}, we adopt a much lower value of $\eta = 1$ as larger values of $\eta$ will only raise the threshold necessary for the strings to be in virial equilibrium.} We find that the average predicted virial mass of the strings is $\approx 2\times 10^{6} \; \rm M_{\sun}$. We approximate the observed mass of each string by counting the number of members and assuming an average stellar mass of $\rm 0.61 \; M_\sun$ based on the Initial Mass Function from \citet{Maschberger_2013} \citep[see also e.g.][]{Kuhn_2019}. We find a typical observed mass of $\rm M_{observed} = 134 \; M_\sun$, meaning that strings on average require $ > 10^{4} \times$ larger masses than their observed masses to be in virial equilibrium. Even assuming a very poor completeness fraction, all strings are gravitationally unbound, counter to the claim in KC19 that strings are ``weakly bound."

While the unbound state of the strings does not in itself imply that strings are unphysical, it does provide constraints on their predicted lifetimes. If a string is gravitationally unbound, it should disperse on roughly a crossing time, $\rm t_{cross}$: 

\begin{equation}
t_{dispersal} \approx t_{cross} \approx \frac{r_{hm}}{\sigma_{V_r}}    
\end{equation}

\begin{figure*}
    \centering
    \includegraphics[width=0.75\textwidth]{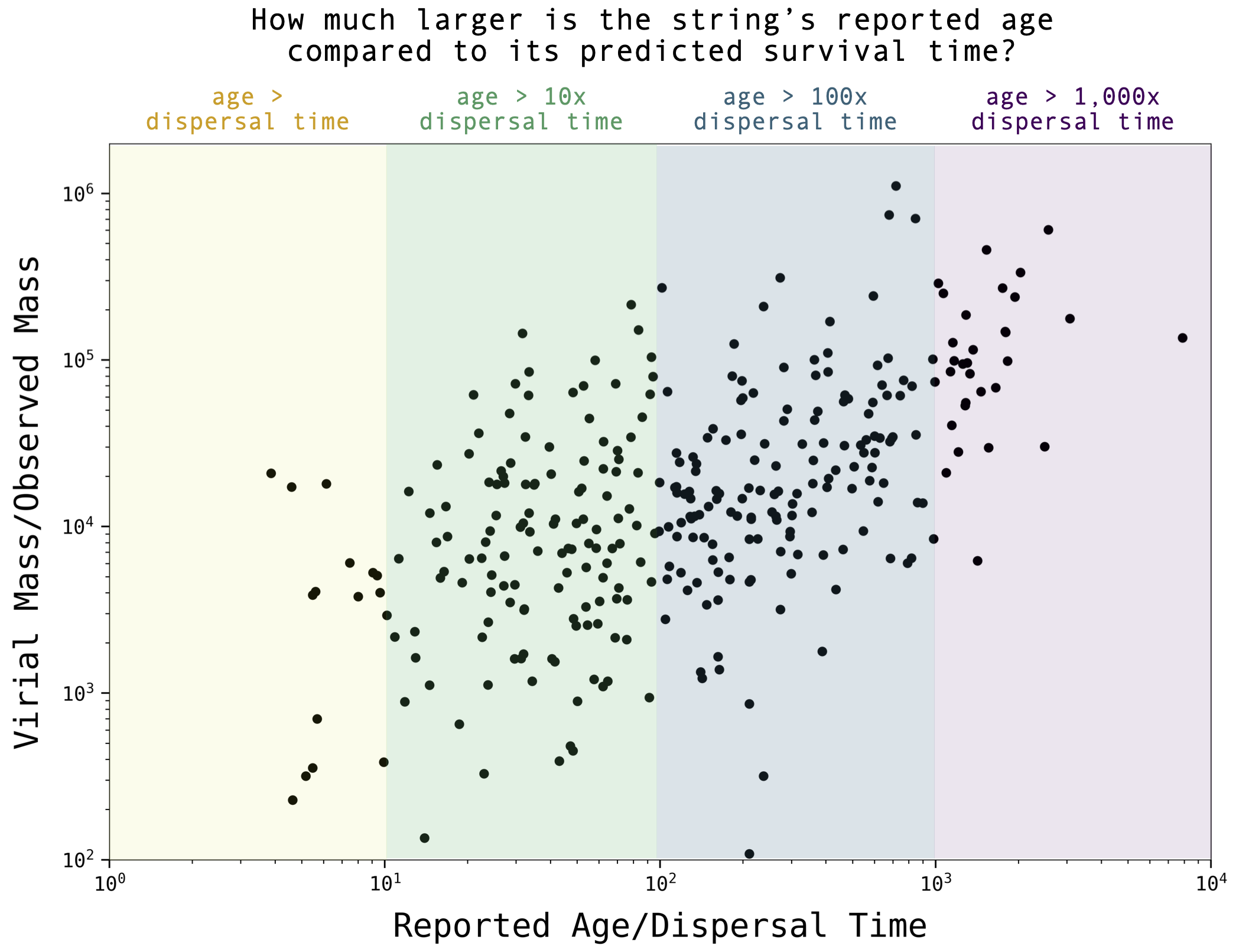}
    \caption{The virial state of strings versus the discrepancy between the strings' reported and predicted lifetimes. The vertical axis shows the ratio of the predicted virial masses of the strings (based on their observed radial velocity dispersions) over their observed masses. Every string in the sample is gravitationally unbound, requiring on average of $2\times10^{4}$ times larger mass than observed to be in virial equilibrium. Given their unbound state, the dispersal time of the strings should be roughly the crossing time. As shown on the horizontal axis, we find that the reported ages of the strings are orders of magnitude larger than the dispersal time, meaning that strings should have dispersed in a small fraction of their reported lifetimes and are thus inconsistent with being physical structures. 
    \label{fig:lifetimes}}
\end{figure*}

We find a median predicted dispersal time for the strings of only 2 Myr. Since KC19 determine ages of between 4 Myr and 9 Gyr for the strings, the strings' reported ages are on average $126\times$ larger than their dispersal times. In Figure \ref{fig:lifetimes} we plot $\frac{\rm M_{virial}}{\rm M_{observed}}$ as a function of $\rm \frac{Reported \; Age}{Dipsersal \; Time}$. Since the strings should have already dispersed in a fraction of their reported lifetimes, their kinematics are likewise inconsistent with the claim that strings are co-eval, physical entities whose members share a common origin.

\subsection{The Chemical Homogeneity of Stellar Strings} \label{subsec:chemical}
\begin{figure*}
    \centering
    \includegraphics[width=1.0\textwidth]{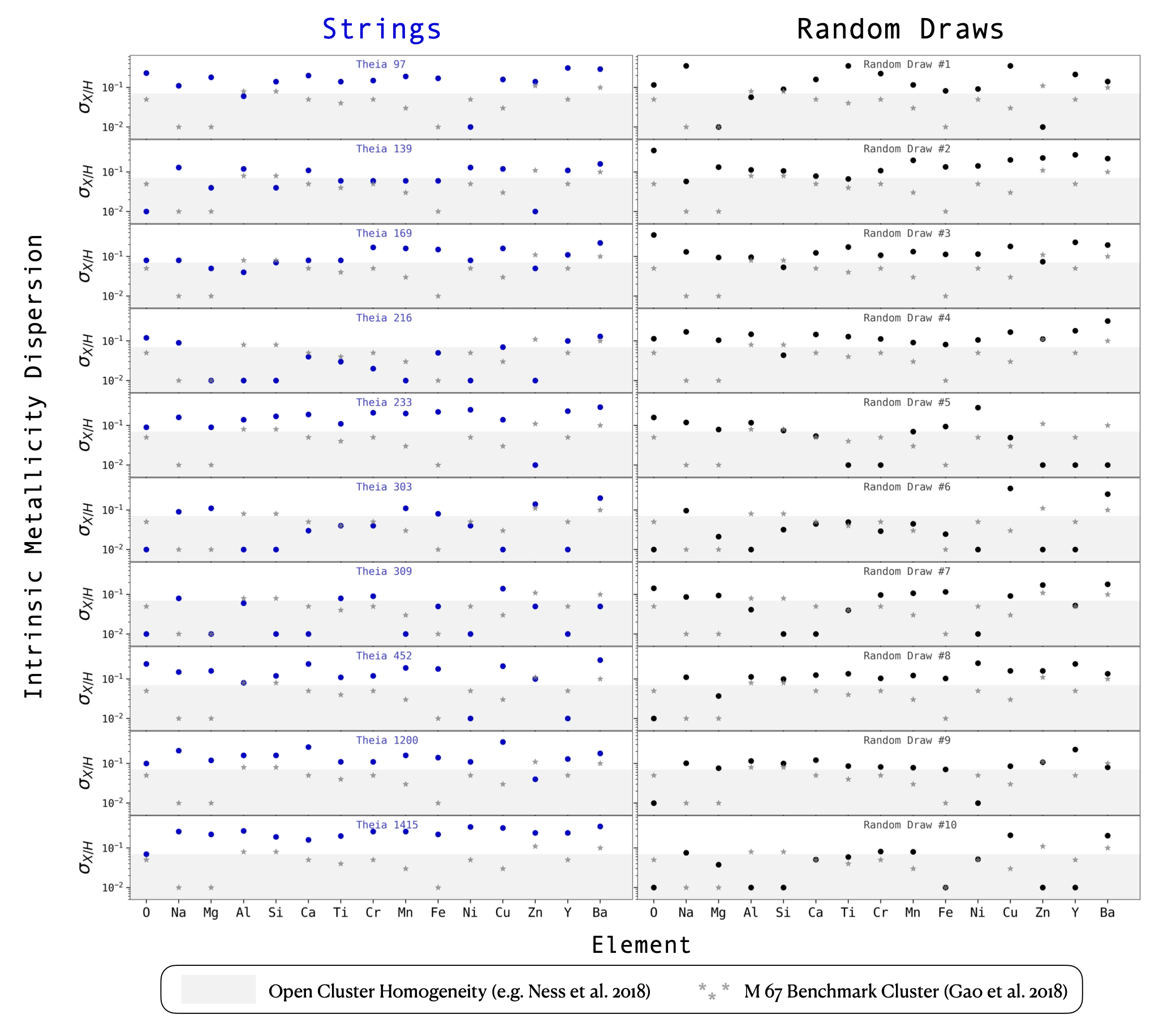}
    \caption{Intrinsic abundance variations across fifteen different elements for the collection of ten strings from MHM22 (blue points, left) and ten random sub-samples of stars drawn from unrelated open clusters in the Milky Way (black points, right). The grey asterisks in each panel show the intrinsic abundance variations for the benchmark cluster M67 \citep{Gao_2018, Manea_2022}, while the gray shaded region marks the zone of reported open cluster dispersions \citep[e.g.][]{Ness_2018}. By drawing random samples of stars from unrelated clusters, we are able to reproduce the typical chemical homogeneity of a majority of the strings, indicating that the strings' abundance variations are not discriminatory enough to favor a common physical origin. An interactive version of this figure showing hundreds of random draws, rather than just ten, is available online \href{https://faun.rc.fas.harvard.edu/czucker/Paper\_Figures/Figure5\_strings\_vs\_random.html}{here}. \label{fig:metallicity}}
\end{figure*}

We perform a final test to determine the physicality of the stellar string members by examining the uniformity in their chemical composition with respect to open cluster members. If a set of stars is born within the same parental molecular gas structure, they should be chemically homogeneous \citep[e.g.][]{Feng_2014}. To examine the chemical homogeneity of the strings, we build off the study of \citet[][MHM22]{Manea_2022}, who leverage GALAH data \citep{GALAH_DR3} to characterize the intrinsic chemical dispersion ($\rm \sigma_{[X/H]}$) of a sample of ten strings. MHM22 fit the following likelihood function assuming that the chemical profile of each string is Gaussian with some mean abundance $\rm \mu_{[X/H]}$ and intrinsic dispersion $\rm \sigma_{[X/H]}$:

\begin{equation}
    \mathcal{L} = \prod_i^N \rm{exp} \left [ \frac{-(x_i - \mu_{[X/H]})^2} {(2(\sigma_{[X/H]}^2 + \delta_i^2)} \right ] \times \frac{1}{\sqrt{2\pi(\sigma_{[X/H]}^2 + \delta_i^2)}}
    \label{eq:likelihood}
\end{equation}

where $x_i$ and $\sigma_i$ are the GALAH mean abundance and its reported uncertainty for the $i$th star in the string in a given element $\rm X$. MHM22 characterize the intrinsic chemical dispersion across a range of elements with a sample size between 7 and 19 stars per string. In the left panel of Figure \ref{fig:metallicity}, we reproduce the original results of MHM22 (see their Figure 4), showing the intrinsic dispersion $\rm \sigma_{[X/H]}$ for each of the ten strings. MHM22 find that all but one of the strings is more homogeneous than their local field stars, with half of the sample as homogeneous as the well-studied open cluster M67 in several elements \citep{Gao_2018}. 

To test whether it is possible for a string to appear chemically homogeneous in several elements \textit{without} being co-eval, we draw random sub-samples of stars from the \citet{Spina_2021} catalog, who curate a sample of open cluster members detected in GALAH. Specifically, following MHM22, we draw between 7 and 19 per sub-sample and restrict to open clusters which span the same broad age range for the strings considered in MHM22 (7.52 $<$ log(Age) $<$ 9.23), have a detection in GALAH, and high membership probability $p > 0.75$. These sub-samples consist of stars that do \textit{belong to well-studied open clusters}, but each sub-sample is drawn from many clusters that are \textit{unrelated}. We then fit the same likelihood function as MHM22, repeating this procedure over many trials.  Since we argue strings are likely agglomerations of unrelated open clusters and other dynamically cold field stars, this experiment provides a more direct comparison point to interpret apparent homogeneity found in MHM22. 

As seen in Figure \ref{fig:metallicity}, we can match the chemical homogeneity of the strings with the random draws, which we attribute to two causes. First, the uncertainties on the GALAH abundance variations for an individual star $i$ (i.e. $\sigma_i$ in Equation \ref{eq:likelihood}) are similar to the intrinsic abundance variation of a cluster like M67 ($\sigma_i \approx \sigma_{[X/H]} \approx 0.1$ dex). Since Equation \ref{eq:likelihood} is designed to capture the intrinsic uncertainty by modeling the observational error, any overestimation of the error in GALAH can lead to unrealistically small estimates for the intrinsic scatter when the errors are large. And second, these random sub-samples can appear more homogeneous than field stars simply by virtue of the stars being members of open clusters, even if these clusters are physically unrelated.\footnote{The local field star sample from \citet{Manea_2022} (showing poorer chemical homogeneity than the strings) also likely includes some thick disk stars, which will have a wider metallicity dispersion than the thin disk string stars selected by KC19.} 

Our results are consistent with the scenario that, while these strings may sometimes contain real clusters, their abundance patterns are not discriminatory enough to favor a scenario where members of the string have the same origin, versus a range of origins in potentially real, yet physically unrelated, spatial sub-structures. The chemical homogeneity of the strings found in MHM22 does not show them to be co-eval.

\section{Discussion} \label{discussion}
Through a spatial, kinematic, and chemical re-analysis of their stellar membership, we have shown that strings are inconsistent with being co-eval stellar structures with a common physical origin. KC19 select all 328 strings through a manual assembly process, stitching together stellar groups by hand and visually confirming kinematic and spatial coherence by eye. Our work underlies the need for a more systematic, \textit{reproducible} selection process when declaring the existence of a new type of stellar structure: not only should these structures remain spatially coherent and continuous when viewed in true 3D physical space, but their radial velocity dispersions should also be significantly smaller than measured for the Galactic field. In the solar neighborhood, the age-velocity dispersion relation (describing how the velocity dispersion of stars appears to increase with age due to dynamical heating) shows typical vertical velocity dispersions of $\rm \approx 5 \; km \; s^{-1}$ for stars $<$ 1 Gyr, about $3\times$ smaller than the typical string radial velocity dispersion of $\rm 16 \; km \; s^{-1}$ \citep[see][]{Bird_2021, Casagrande_2011}. The age-velocity dispersion relation has also been explored in 3D using open clusters. Specifically, for open clusters in the same age range as the typical string (150 - 250 Myr), \citet{Tarricq_2021} find a 3D velocity dispersion of $\rm 13 \; km \; s^{-1}$, meaning that the velocity dispersion \textit{over} a sample of many open clusters is less than the typical radial velocity dispersion \textit{within} an individual string in KC19 \citep[see Figure 11, Table 3 in][]{Tarricq_2021}.

While we argue against the physicality of strings in KC19, several other studies in the \textit{Gaia} era present compelling evidence for filamentary stellar distributions identified though more reproducible selection algorithms, some of which are summarized in Table \ref{tab:litcomp}. \citet{Meingast_2019} identify an extended 400+ pc long, 2000 $\rm M_\odot$ stream in the disk called Meingast-1 \citep[also known as Pisces Eridanus; e.g.][]{Hawkins_2020} through a wavelet decomposition of the 3D velocity space distribution of nearby stars. Not only do \citet{Meingast_2019} find that the stream is spatially continuous in 3D space but they also find a 3D velocity dispersion of $\rm 1.3 \; km \;s^{-1}$. Similarly, \citet{Meingast_2021} present a new method for identifying highly extended coronae around ten nearby open clusters. The \citet{Meingast_2021} technique accounts for projection effects in proper motion space in an automated way \citep[inspired by the ``convergent point technique"; c.f.][]{van_Leeuwen_2009} before deconvolving the spatial distribution with a Gaussian mixture model to mitigate \textit{Gaia} measurement errors. The coronae are likewise validated via their 3D space motions, showing typical 3D velocity dispersions of $\rm 1.4 \; km \; s^{-1}$.

Both the extended coronae and the Meingast-1 stream have velocity dispersions on par with open clusters. Using a sample of a few hundred nearby open clusters, \citet{Soubiran_2018} find typical intra-cluster radial velocity dispersion of $\rm 1.0-1.5 \; km \; s^{-1}$. Only four strings in KC19 have radial velocity dispersions $\rm < 5 \; km \; s^{-1}$ (Theia 127, 161, 605, and 998), while $\approx$ 90\% of the open clusters do \citep{Soubiran_2018}, as well as 100\% of the newly identified extended structures in \citet{Meingast_2019} and \citet{Meingast_2021}. The unphysical nature of the strings is not due to their claimed unique filamentary morphologies, but rather their lack of true 3D kinematic and spatial coherence stemming from limitations in the manual assembly process. Since we argue that some strings can be composed of open clusters and other dynamically cold field stars, dedicated follow-up studies \citep[e.g.][]{Andrews_2022} would be needed to characterize the extent to which individual strings may contain physically relevant sub-structure. 

\section{Conclusions} \label{conclusions}

We investigate the spatial, dynamical, and chemical composition of stellar strings, a proposed collection of highly extended filamentary stellar structures identified in KC19 by manually linking stellar groups with similar 5D properties ($l$, $b$, parallax $\pi$, and proper motions). Our conclusions are as follows:

\begin{enumerate}

\item Using updated constraints on the distances to stellar string members from \textit{Gaia} DR3, we find that the 3D spatial dispersion of stars around the string spine does not improve over \textit{Gaia} DR2: the average percentage that stars move closer to their respective string spines is consistent with zero, despite the signal to noise on the parallax measurements per string increasing by 20\% - 120\%. Real structures should tighten with higher fidelity distance measurements.

\item The average dispersion in the radial velocity of the strings is $\rm 16 \; km \; s^{-1}$, about fifteen times larger than the typical radial velocity dispersion both of open clusters in \textit{Gaia} \citep{Soubiran_2018} and in other catalogs of extended stellar structures \citep[e.g. stellar ``streams" in the disk from][]{Meingast_2019, Meingast_2021}.

\item Given the radial velocity dispersions, the virial masses of the strings are on average $> 10^{4}\times$ larger than their observed masses. Even assuming very low completeness fractions, all strings are gravitationally unbound.

\item Given their unbound state, the strings should disperse on roughly a crossing time, which we estimate to be typically 2 Myr, while the ages of the strings from KC19 range from 4 Myr to 9 Gyr. Thus the strings should not exist based on their predicted dynamical lifetimes, and should have dispersed in $<$ 1\% of their reported ages, on average. 

\item Using complementary constraints on stellar chemical abundances from GALAH DR3 \citep{GALAH_DR3}, we compare the intrinsic abundance dispersion of the strings found in MHM22 to a random sample of stars drawn from physically unrelated open clusters.  We find that the chemical homogeneity of the strings is similar to the chemical homogeneity seen in random stellar draws across clusters. 

\item The combined spatial, dynamical, and chemical evidence rules out the scenario that stars within a typical string were born at the same time within the same parental molecular gas structure. However, some subset of the stars within a string may still be co-eval, as many of these strings contain real clusters that have been linked together to form the larger string-like structure. 

\end{enumerate}

Ultimately, by evoking simple spatial and dynamical arguments, our work provides a straightforward, yet discerning, lens to evaluate the fidelity of newfound classes of objects, which should be considered when declaring the existence of new co-eval stellar structures in the \textit{Gaia} era.

\begin{acknowledgments}
The authors would like to thank the anonymous referee for their thorough and prompt review of our manuscript. CZ acknowledges that support for this work was provided by NASA through the NASA Hubble Fellowship grant \#HST-HF2-51498.001 awarded by the Space Telescope Science Institute (STScI), which is operated by the Association of Universities for Research in Astronomy, Inc., for NASA, under contract NAS5-26555. SL acknowledges support from NSF grant AST2109234 and HST-AR-16624 from STScI. The authors would like to thank Keith Hawkins, Catherine Manea, Kevin Covey, Marina Kounkel, Stefan Meingast, João Alves, and Alyssa Goodman for helpful comments that contributed to this work. 

\end{acknowledgments}


\bibliography{sample631}{}
\bibliographystyle{aasjournal}

\include{Table2}

\appendix

\section{Gallery of Stellar Strings} \label{gallery_appendix}

\figsetstart
\figsetnum{A1}
\figsettitle{Stellar Strings Gallery}

\figsetgrpstart
\figsetgrpnum{A1.1}
\figsetgrptitle{Page 1}
\figsetplot{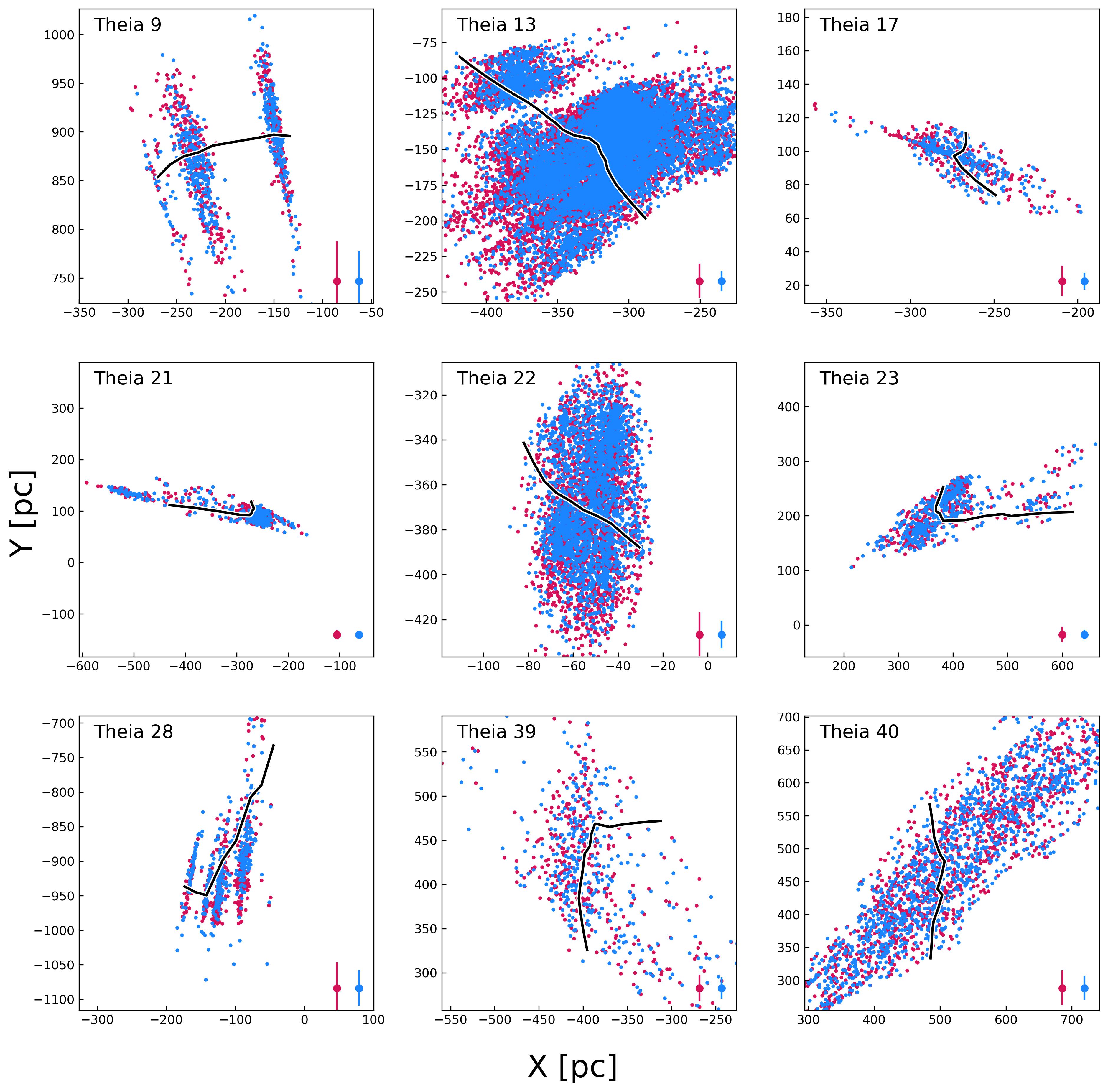}
\figsetgrpnote{Top-down view of stellar strings, with each panel showing the $Gaia$ DR2 (red) and $Gaia$ DR3 (blue) stellar distribution of stars in the string, alongside the ``spine" shown in black. The lower right hand corner of each panel shows the typical distance errors for stars in the string. }
\figsetgrpend

\figsetend

\begin{figure}[h!]
\figurenum{A1}
\plotone{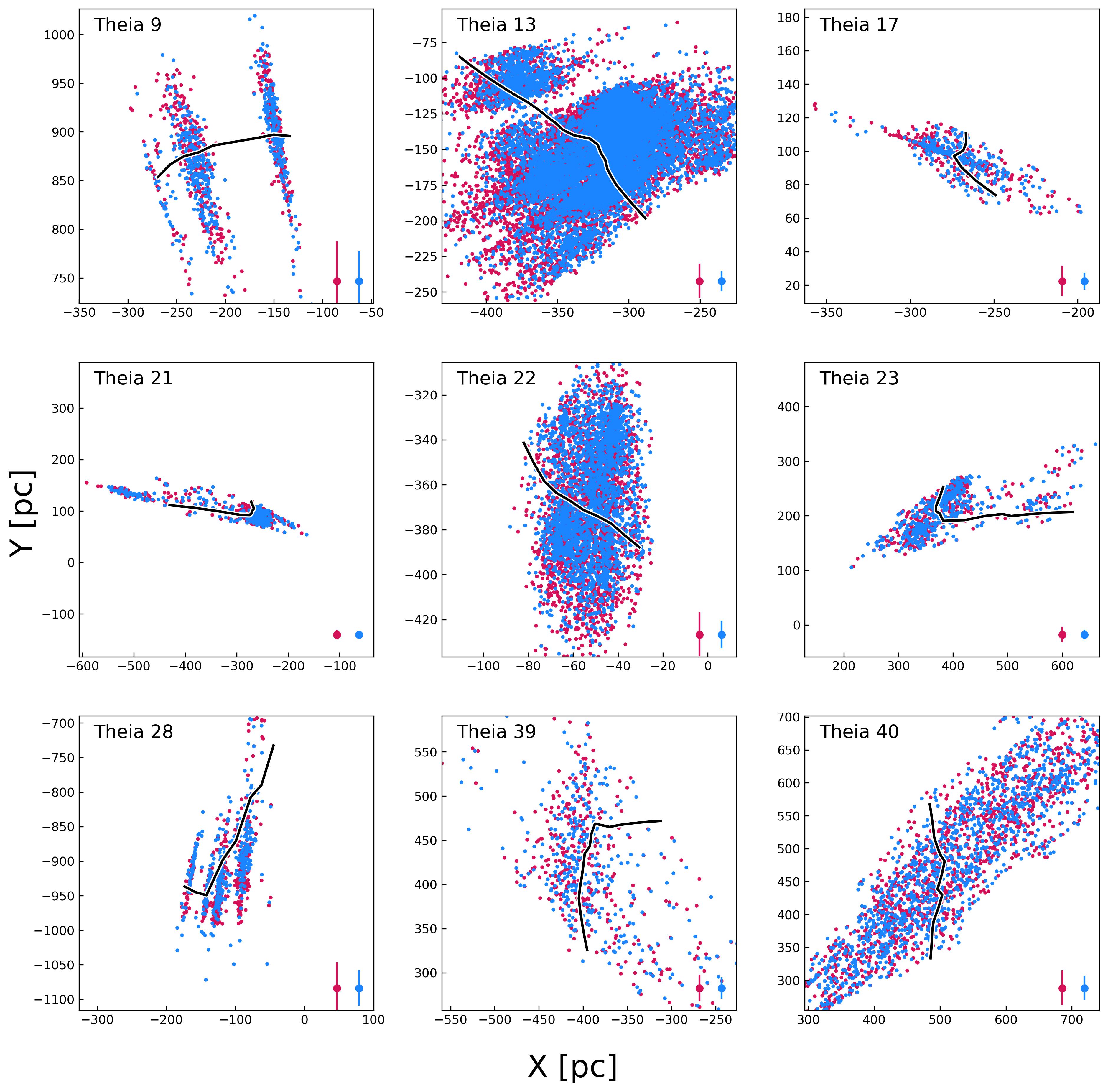}
\caption{Top-down view of stellar strings, with each panel showing the $Gaia$ DR2 (red) and $Gaia$ DR3 (blue) stellar distribution of stars in the string, alongside the ``spine" shown in black. The lower right hand corner of each panel shows the typical distance errors for stars in the string. An online version of this figure set is available \href{https://faun.rc.fas.harvard.edu/czucker/Paper_Figures/String_Gallery_Interactive.html}{here}. \label{fig:figset}}
\end{figure}

\end{document}

%% file: Table2.tex
\tabletypesize{\tiny}
\setlength{\tabcolsep}{7pt}
\startlongtable
\begin{deluxetable*}{cccccccccccc}
\tablecaption{Spatial and Dynamical Properties of Stellar Strings\label{tab:summary} }
\colnumbers
\tablehead{\colhead{Theia} & \colhead{DR2 Offset} & \colhead{DR2 S/N} & \colhead{DR3 Offset} & \colhead{DR3 S/N} & \colhead{$\sigma_{V_r}$} & \colhead{$\rm M_{virial}$} & \colhead{$\rm M_{observed}$} & \colhead{$\rm \frac{M_{virial}}{M_{observed}}$} & \colhead{$\rm t_{dispersal}$} & \colhead{Age} & \colhead{$\rm \frac{Age}{t_{dispersal}}$} \\
\colhead{} & \colhead{pc} & \colhead{} & \colhead{pc} & \colhead{} & \colhead{$\rm km\; s^{-1}$} & \colhead{$\rm M_\odot$} & \colhead{$\rm M_\odot$} & \colhead{} & \colhead{Myr} & \colhead{Myr} & \colhead{}}
\startdata
9 & 35 & 22 & 28 & 30 & 29.2 & 5.7e+06 & 327 & 1.7e+04 & 1.0 & 4 & 4 \\
13 & 29 & 31 & 27 & 52 & 17.8 & 2.1e+06 & 6457 & 3.2e+02 & 1.5 & 7 & 5 \\
17 & 16 & 32 & 15 & 58 & 16.7 & 9.9e+05 & 194 & 5.1e+03 & 0.9 & 8 & 9 \\
21 & 24 & 32 & 24 & 55 & 15.5 & 1.4e+06 & 225 & 6.0e+03 & 1.5 & 11 & 7 \\
22 & 23 & 38 & 22 & 60 & 10.2 & 5.6e+05 & 1563 & 3.6e+02 & 2.2 & 12 & 5 \\
23 & 31 & 30 & 31 & 48 & 53.2 & 2.1e+07 & 331 & 6.2e+04 & 0.6 & 12 & 20 \\
28 & 28 & 26 & 27 & 34 & 12.7 & 1.0e+06 & 247 & 4.1e+03 & 2.1 & 11 & 5 \\
39 & 33 & 39 & 31 & 47 & 10.6 & 8.2e+05 & 212 & 3.9e+03 & 2.9 & 15 & 5 \\
40 & 87 & 27 & 91 & 38 & 27.4 & 1.6e+07 & 763 & 2.1e+04 & 3.3 & 12 & 3 \\
41 & 28 & 18 & 25 & 25 & 28.6 & 4.8e+06 & 400 & 1.2e+04 & 0.9 & 12 & 14 \\
43 & 10 & 97 & 10 & 214 & 8.9 & 2.0e+05 & 1471 & 1.4e+02 & 1.2 & 16 & 13 \\
44 & 10 & 97 & 10 & 206 & 7.7 & 1.4e+05 & 161 & 8.9e+02 & 1.3 & 15 & 11 \\
45 & 15 & 84 & 15 & 195 & 17.3 & 1.0e+06 & 1600 & 6.5e+02 & 0.8 & 15 & 18 \\
53 & 4 & 72 & 4 & 152 & 11.1 & 1.1e+05 & 26 & 4.3e+03 & 0.4 & 25 & 70 \\
55 & 44 & 42 & 43 & 70 & 17.3 & 3.0e+06 & 1017 & 2.9e+03 & 2.4 & 24 & 10 \\
56 & 45 & 22 & 85 & 28 & 22.5 & 1.0e+07 & 553 & 1.8e+04 & 3.7 & 22 & 6 \\
57 & 37 & 29 & 37 & 44 & 9.9 & 8.5e+05 & 1209 & 7.0e+02 & 3.7 & 21 & 5 \\
58 & 43 & 36 & 43 & 51 & 15.5 & 2.5e+06 & 463 & 5.3e+03 & 2.8 & 25 & 9 \\
72 & 14 & 39 & 14 & 64 & 34.9 & 4.0e+06 & 116 & 3.4e+04 & 0.4 & 31 & 78 \\
73 & 33 & 43 & 31 & 66 & 16.9 & 2.1e+06 & 1873 & 1.1e+03 & 1.8 & 26 & 14 \\
74 & 38 & 47 & 37 & 71 & 5.4 & 2.6e+05 & 1141 & 2.3e+02 & 6.8 & 31 & 4 \\
75 & 34 & 50 & 33 & 71 & 14.5 & 1.6e+06 & 997 & 1.6e+03 & 2.3 & 29 & 12 \\
77 & 18 & 31 & 16 & 47 & 20.3 & 1.6e+06 & 506 & 3.2e+03 & 0.8 & 26 & 32 \\
79 & 32 & 47 & 28 & 62 & 18.3 & 2.2e+06 & 256 & 8.7e+03 & 1.5 & 25 & 16 \\
80 & 21 & 31 & 23 & 44 & 11.0 & 6.5e+05 & 278 & 2.3e+03 & 2.1 & 26 & 12 \\
81 & 26 & 26 & 31 & 36 & 11.2 & 9.1e+05 & 226 & 4.0e+03 & 2.7 & 26 & 9 \\
86 & 22 & 35 & 21 & 53 & 9.0 & 4.1e+05 & 186 & 2.2e+03 & 2.4 & 25 & 10 \\
87 & 49 & 36 & 38 & 53 & 11.5 & 1.2e+06 & 312 & 3.8e+03 & 3.3 & 26 & 7 \\
91 & 44 & 39 & 38 & 53 & 15.7 & 2.2e+06 & 340 & 6.4e+03 & 2.4 & 26 & 11 \\
92 & 9 & 114 & 9 & 205 & 5.3 & 6.1e+04 & 818 & 7.4e+01 & 1.7 & 35 & 20 \\
95 & 25 & 74 & 24 & 113 & 16.1 & 1.5e+06 & 685 & 2.2e+03 & 1.5 & 33 & 22 \\
96 & 21 & 46 & 20 & 70 & 13.5 & 8.7e+05 & 775 & 1.1e+03 & 1.5 & 35 & 23 \\
97 & 25 & 43 & 24 & 61 & 21.5 & 2.7e+06 & 595 & 4.5e+03 & 1.1 & 33 & 29 \\
98 & 31 & 52 & 30 & 75 & 12.1 & 1.0e+06 & 211 & 4.9e+03 & 2.4 & 38 & 15 \\
99 & 44 & 46 & 45 & 65 & 23.0 & 5.6e+06 & 427 & 1.3e+04 & 1.9 & 32 & 16 \\
108 & 25 & 44 & 24 & 63 & 13.9 & 1.1e+06 & 240 & 4.6e+03 & 1.7 & 33 & 19 \\
111 & 41 & 35 & 41 & 47 & 15.3 & 2.2e+06 & 138 & 1.6e+04 & 2.6 & 32 & 12 \\
113 & 11 & 201 & 11 & 396 & 16.2 & 6.7e+05 & 111 & 6.0e+03 & 0.7 & 43 & 64 \\
115 & 27 & 95 & 27 & 156 & 14.6 & 1.4e+06 & 339 & 4.0e+03 & 1.8 & 44 & 24 \\
116 & 26 & 56 & 26 & 92 & 24.3 & 3.6e+06 & 681 & 5.3e+03 & 1.1 & 48 & 45 \\
118 & 23 & 50 & 23 & 80 & 15.7 & 1.3e+06 & 1133 & 1.2e+03 & 1.5 & 50 & 34 \\
120 & 21 & 47 & 20 & 67 & 13.6 & 8.7e+05 & 508 & 1.7e+03 & 1.4 & 46 & 31 \\
125 & 35 & 33 & 34 & 50 & 22.7 & 4.1e+06 & 340 & 1.2e+04 & 1.5 & 49 & 33 \\
127 & 25 & 36 & 22 & 50 & 4.6 & 1.1e+05 & 290 & 3.9e+02 & 4.9 & 48 & 9 \\
133 & 19 & 96 & 18 & 160 & 21.0 & 1.9e+06 & 1612 & 1.2e+03 & 0.9 & 55 & 64 \\
136 & 8 & 56 & 7 & 83 & 7.3 & 9.2e+04 & 103 & 8.9e+02 & 1.0 & 50 & 50 \\
138 & 13 & 38 & 13 & 57 & 9.3 & 2.6e+05 & 171 & 1.5e+03 & 1.4 & 57 & 41 \\
139 & 29 & 40 & 31 & 67 & 19.3 & 2.7e+06 & 78 & 3.5e+04 & 1.6 & 51 & 32 \\
143 & 22 & 54 & 21 & 77 & 8.5 & 3.6e+05 & 1081 & 3.3e+02 & 2.4 & 55 & 22 \\
149 & 42 & 42 & 43 & 59 & 30.0 & 9.2e+06 & 442 & 2.1e+04 & 1.4 & 57 & 40 \\
150 & 31 & 44 & 33 & 56 & 14.2 & 1.6e+06 & 195 & 8.1e+03 & 2.3 & 53 & 23 \\
153 & 34 & 35 & 36 & 45 & 16.1 & 2.2e+06 & 186 & 1.2e+04 & 2.2 & 56 & 25 \\
157 & 54 & 32 & 49 & 39 & 29.9 & 1.0e+07 & 168 & 6.1e+04 & 1.6 & 54 & 33 \\
158 & 28 & 33 & 30 & 46 & 15.5 & 1.7e+06 & 162 & 1.0e+04 & 1.9 & 60 & 31 \\
160 & 16 & 71 & 16 & 113 & 17.1 & 1.1e+06 & 183 & 6.1e+03 & 0.9 & 79 & 84 \\
161 & 8 & 64 & 8 & 94 & 4.9 & 4.5e+04 & 93 & 4.8e+02 & 1.6 & 75 & 47 \\
163 & 21 & 67 & 21 & 94 & 18.3 & 1.6e+06 & 442 & 3.7e+03 & 1.1 & 78 & 69 \\
164 & 20 & 68 & 20 & 100 & 8.1 & 3.1e+05 & 194 & 1.6e+03 & 2.5 & 78 & 31 \\
165 & 42 & 51 & 43 & 74 & 18.8 & 3.5e+06 & 1116 & 3.2e+03 & 2.2 & 71 & 32 \\
166 & 30 & 45 & 30 & 62 & 8.9 & 5.6e+05 & 87 & 6.5e+03 & 3.4 & 75 & 22 \\
169 & 31 & 45 & 34 & 63 & 10.5 & 8.9e+05 & 173 & 5.1e+03 & 3.2 & 78 & 24 \\
174 & 61 & 42 & 63 & 52 & 17.0 & 4.3e+06 & 156 & 2.7e+04 & 3.7 & 73 & 20 \\
182 & 34 & 28 & 38 & 42 & 43.4 & 1.7e+07 & 77 & 2.2e+05 & 0.9 & 67 & 78 \\
189 & 50 & 30 & 50 & 47 & 17.7 & 3.7e+06 & 205 & 1.8e+04 & 2.8 & 71 & 25 \\
190 & 92 & 22 & 107 & 35 & 45.8 & 5.2e+07 & 361 & 1.4e+05 & 2.3 & 72 & 31 \\
195 & 87 & 21 & 92 & 34 & 36.1 & 2.8e+07 & 584 & 4.8e+04 & 2.5 & 70 & 28 \\
205 & 26 & 30 & 32 & 42 & 15.7 & 1.9e+06 & 105 & 1.8e+04 & 2.0 & 70 & 34 \\
208 & 11 & 217 & 11 & 478 & 19.8 & 1.1e+06 & 67 & 1.6e+04 & 0.6 & 94 & 163 \\
213 & 31 & 74 & 32 & 103 & 8.9 & 6.0e+05 & 90 & 6.7e+03 & 3.6 & 97 & 27 \\
214 & 19 & 57 & 19 & 82 & 18.6 & 1.6e+06 & 170 & 9.1e+03 & 1.0 & 96 & 95 \\
215 & 29 & 84 & 28 & 121 & 34.4 & 7.7e+06 & 278 & 2.8e+04 & 0.8 & 91 & 114 \\
216 & 13 & 61 & 13 & 84 & 5.8 & 1.1e+05 & 269 & 3.9e+02 & 2.3 & 97 & 42 \\
219 & 25 & 57 & 24 & 88 & 17.8 & 1.8e+06 & 160 & 1.1e+04 & 1.3 & 94 & 70 \\
228 & 24 & 45 & 29 & 58 & 8.2 & 4.6e+05 & 104 & 4.4e+03 & 3.6 & 96 & 27 \\
233 & 58 & 53 & 60 & 73 & 12.0 & 2.0e+06 & 376 & 5.4e+03 & 4.9 & 80 & 16 \\
236 & 41 & 43 & 43 & 64 & 19.9 & 4.0e+06 & 133 & 3.0e+04 & 2.1 & 84 & 39 \\
237 & 38 & 49 & 40 & 71 & 13.2 & 1.6e+06 & 81 & 2.0e+04 & 3.0 & 79 & 26 \\
244 & 30 & 63 & 27 & 93 & 19.9 & 2.5e+06 & 76 & 3.2e+04 & 1.3 & 82 & 62 \\
245 & 18 & 71 & 18 & 101 & 11.0 & 5.3e+05 & 93 & 5.7e+03 & 1.7 & 90 & 53 \\
246 & 30 & 47 & 32 & 66 & 11.2 & 9.4e+05 & 266 & 3.5e+03 & 2.8 & 80 & 28 \\
248 & 60 & 46 & 58 & 63 & 12.4 & 2.1e+06 & 330 & 6.4e+03 & 4.6 & 93 & 20 \\
250 & 16 & 45 & 15 & 68 & 9.9 & 3.5e+05 & 107 & 3.3e+03 & 1.5 & 81 & 53 \\
252 & 66 & 35 & 81 & 50 & 20.0 & 7.6e+06 & 208 & 3.6e+04 & 4.0 & 87 & 21 \\
253 & 66 & 44 & 69 & 61 & 12.1 & 2.4e+06 & 100 & 2.3e+04 & 5.6 & 87 & 15 \\
254 & 46 & 35 & 44 & 49 & 25.6 & 6.8e+06 & 707 & 9.6e+03 & 1.7 & 99 & 58 \\
259 & 22 & 37 & 25 & 48 & 10.0 & 5.8e+05 & 81 & 7.1e+03 & 2.5 & 88 & 35 \\
261 & 37 & 27 & 38 & 41 & 21.1 & 4.0e+06 & 159 & 2.5e+04 & 1.8 & 93 & 52 \\
262 & 35 & 39 & 38 & 52 & 10.5 & 9.9e+05 & 104 & 9.4e+03 & 3.5 & 85 & 24 \\
263 & 38 & 41 & 33 & 59 & 25.3 & 5.0e+06 & 197 & 2.5e+04 & 1.3 & 92 & 70 \\
275 & 72 & 25 & 60 & 42 & 21.5 & 6.5e+06 & 90 & 7.2e+04 & 2.8 & 82 & 29 \\
282 & 55 & 25 & 49 & 40 & 17.9 & 3.6e+06 & 203 & 1.8e+04 & 2.7 & 86 & 32 \\
285 & 55 & 29 & 60 & 38 & 19.1 & 5.1e+06 & 238 & 2.2e+04 & 3.1 & 82 & 26 \\
291 & 57 & 22 & 67 & 35 & 36.7 & 2.1e+07 & 328 & 6.4e+04 & 1.8 & 86 & 48 \\
294 & 34 & 24 & 33 & 39 & 12.3 & 1.2e+06 & 731 & 1.6e+03 & 2.7 & 79 & 29 \\
301 & 24 & 125 & 24 & 191 & 13.0 & 9.7e+05 & 801 & 1.2e+03 & 1.9 & 106 & 57 \\
303 & 19 & 74 & 18 & 129 & 25.1 & 2.7e+06 & 204 & 1.3e+04 & 0.7 & 107 & 149 \\
305 & 9 & 78 & 7 & 110 & 15.2 & 3.9e+05 & 25 & 1.6e+04 & 0.5 & 122 & 259 \\
306 & 19 & 45 & 18 & 61 & 9.1 & 3.5e+05 & 43 & 7.9e+03 & 1.9 & 106 & 55 \\
308 & 14 & 66 & 14 & 97 & 16.3 & 9.0e+05 & 78 & 1.1e+04 & 0.9 & 110 & 128 \\
309 & 24 & 58 & 24 & 78 & 12.8 & 9.3e+05 & 355 & 2.6e+03 & 1.8 & 109 & 59 \\
310 & 26 & 57 & 26 & 76 & 15.8 & 1.5e+06 & 101 & 1.5e+04 & 1.7 & 105 & 64 \\
311 & 38 & 50 & 36 & 72 & 14.7 & 1.8e+06 & 264 & 6.9e+03 & 2.4 & 106 & 43 \\
312 & 18 & 42 & 18 & 61 & 13.7 & 8.0e+05 & 62 & 1.3e+04 & 1.3 & 100 & 77 \\
316 & 40 & 30 & 46 & 42 & 39.6 & 1.7e+07 & 211 & 8.0e+04 & 1.1 & 107 & 94 \\
325 & 58 & 30 & 57 & 48 & 11.5 & 1.8e+06 & 659 & 2.7e+03 & 4.9 & 116 & 23 \\
326 & 37 & 30 & 40 & 45 & 18.1 & 3.0e+06 & 68 & 4.5e+04 & 2.2 & 119 & 55 \\
327 & 42 & 44 & 39 & 60 & 19.1 & 3.3e+06 & 206 & 1.6e+04 & 2.0 & 101 & 50 \\
329 & 43 & 31 & 44 & 50 & 29.5 & 9.0e+06 & 125 & 7.2e+04 & 1.5 & 101 & 68 \\
331 & 23 & 44 & 20 & 64 & 12.8 & 8.0e+05 & 371 & 2.2e+03 & 1.6 & 109 & 68 \\
334 & 34 & 28 & 26 & 44 & 16.2 & 1.6e+06 & 217 & 7.4e+03 & 1.6 & 106 & 66 \\
339 & 41 & 33 & 40 & 47 & 11.6 & 1.3e+06 & 134 & 9.3e+03 & 3.4 & 114 & 33 \\
342 & 29 & 30 & 29 & 45 & 12.9 & 1.1e+06 & 152 & 7.4e+03 & 2.2 & 101 & 46 \\
344 & 39 & 28 & 42 & 42 & 31.4 & 9.8e+06 & 64 & 1.5e+05 & 1.3 & 111 & 83 \\
348 & 43 & 27 & 43 & 44 & 17.6 & 3.2e+06 & 431 & 7.3e+03 & 2.4 & 116 & 47 \\
349 & 51 & 36 & 49 & 49 & 29.1 & 9.7e+06 & 434 & 2.2e+04 & 1.7 & 102 & 62 \\
353 & 73 & 28 & 81 & 40 & 21.5 & 8.8e+06 & 365 & 2.4e+04 & 3.7 & 105 & 28 \\
354 & 86 & 23 & 91 & 36 & 12.9 & 3.5e+06 & 436 & 8.0e+03 & 6.9 & 106 & 15 \\
357 & 60 & 22 & 57 & 30 & 13.4 & 2.4e+06 & 131 & 1.8e+04 & 4.2 & 114 & 27 \\
358 & 68 & 36 & 63 & 51 & 13.7 & 2.8e+06 & 149 & 1.8e+04 & 4.6 & 108 & 23 \\
363 & 48 & 24 & 39 & 33 & 11.2 & 1.1e+06 & 114 & 1.0e+04 & 3.4 & 105 & 31 \\
368 & 9 & 88 & 8 & 155 & 13.8 & 3.7e+05 & 77 & 4.8e+03 & 0.6 & 126 & 213 \\
372 & 17 & 70 & 17 & 123 & 13.3 & 7.3e+05 & 84 & 8.7e+03 & 1.3 & 150 & 115 \\
375 & 15 & 64 & 13 & 112 & 18.2 & 1.1e+06 & 32 & 3.3e+04 & 0.8 & 130 & 173 \\
379 & 30 & 47 & 28 & 75 & 26.7 & 4.8e+06 & 431 & 1.1e+04 & 1.1 & 137 & 129 \\
380 & 24 & 41 & 25 & 66 & 11.8 & 8.4e+05 & 169 & 4.9e+03 & 2.1 & 131 & 62 \\
384 & 56 & 44 & 58 & 56 & 36.3 & 1.8e+07 & 286 & 6.2e+04 & 1.6 & 143 & 91 \\
388 & 30 & 43 & 30 & 59 & 10.6 & 8.1e+05 & 290 & 2.8e+03 & 2.8 & 137 & 48 \\
389 & 35 & 31 & 31 & 47 & 17.9 & 2.4e+06 & 111 & 2.1e+04 & 1.7 & 142 & 82 \\
390 & 34 & 32 & 33 & 51 & 19.9 & 3.1e+06 & 301 & 1.0e+04 & 1.6 & 132 & 82 \\
391 & 46 & 39 & 44 & 58 & 12.7 & 1.7e+06 & 389 & 4.3e+03 & 3.5 & 148 & 42 \\
400 & 81 & 26 & 77 & 40 & 19.9 & 7.2e+06 & 84 & 8.5e+04 & 3.8 & 127 & 33 \\
403 & 57 & 25 & 60 & 39 & 41.6 & 2.5e+07 & 90 & 2.7e+05 & 1.4 & 145 & 101 \\
405 & 72 & 30 & 69 & 44 & 19.8 & 6.3e+06 & 569 & 1.1e+04 & 3.4 & 141 & 41 \\
407 & 25 & 25 & 23 & 44 & 29.8 & 4.8e+06 & 392 & 1.2e+04 & 0.8 & 138 & 180 \\
421 & 67 & 28 & 66 & 37 & 16.2 & 4.1e+06 & 223 & 1.8e+04 & 4.0 & 139 & 35 \\
424 & 6 & 209 & 6 & 394 & 7.2 & 7.9e+04 & 91 & 8.6e+02 & 0.9 & 189 & 210 \\
428 & 13 & 121 & 12 & 174 & 18.4 & 1.0e+06 & 61 & 1.6e+04 & 0.7 & 180 & 268 \\
430 & 11 & 88 & 11 & 139 & 8.2 & 1.8e+05 & 130 & 1.3e+03 & 1.4 & 190 & 140 \\
431 & 20 & 74 & 20 & 119 & 34.4 & 5.7e+06 & 181 & 3.1e+04 & 0.6 & 191 & 328 \\
432 & 13 & 80 & 14 & 121 & 10.8 & 4.0e+05 & 46 & 8.6e+03 & 1.3 & 173 & 131 \\
433 & 33 & 62 & 32 & 90 & 18.3 & 2.5e+06 & 255 & 1.0e+04 & 1.8 & 187 & 107 \\
435 & 15 & 66 & 15 & 106 & 27.7 & 2.8e+06 & 416 & 6.8e+03 & 0.6 & 176 & 315 \\
436 & 22 & 76 & 23 & 107 & 13.2 & 9.5e+05 & 101 & 9.4e+03 & 1.7 & 172 & 99 \\
438 & 22 & 60 & 22 & 98 & 24.4 & 3.1e+06 & 85 & 3.6e+04 & 0.9 & 174 & 196 \\
439 & 24 & 60 & 24 & 82 & 13.6 & 1.0e+06 & 376 & 2.8e+03 & 1.8 & 182 & 104 \\
441 & 31 & 45 & 34 & 71 & 26.8 & 5.8e+06 & 170 & 3.4e+04 & 1.3 & 188 & 148 \\
443 & 26 & 44 & 26 & 58 & 25.6 & 4.0e+06 & 342 & 1.2e+04 & 1.0 & 189 & 190 \\
445 & 28 & 48 & 25 & 64 & 12.5 & 9.5e+05 & 202 & 4.7e+03 & 2.0 & 188 & 92 \\
446 & 25 & 47 & 28 & 63 & 9.2 & 5.7e+05 & 161 & 3.6e+03 & 3.1 & 184 & 60 \\
447 & 35 & 47 & 31 & 69 & 13.3 & 1.3e+06 & 354 & 3.6e+03 & 2.3 & 174 & 75 \\
448 & 41 & 41 & 42 & 60 & 12.7 & 1.6e+06 & 151 & 1.0e+04 & 3.2 & 160 & 49 \\
450 & 37 & 38 & 38 & 52 & 25.3 & 5.8e+06 & 358 & 1.6e+04 & 1.5 & 191 & 127 \\
452 & 44 & 44 & 45 & 62 & 12.8 & 1.7e+06 & 675 & 2.5e+03 & 3.5 & 170 & 49 \\
453 & 31 & 54 & 32 & 75 & 21.7 & 3.5e+06 & 331 & 1.1e+04 & 1.4 & 172 & 119 \\
455 & 53 & 45 & 51 & 58 & 13.7 & 2.2e+06 & 131 & 1.7e+04 & 3.7 & 190 & 51 \\
456 & 25 & 36 & 21 & 58 & 7.9 & 3.1e+05 & 285 & 1.1e+03 & 2.7 & 165 & 61 \\
459 & 59 & 27 & 65 & 42 & 27.0 & 1.1e+07 & 523 & 2.1e+04 & 2.4 & 164 & 69 \\
463 & 54 & 29 & 53 & 44 & 12.0 & 1.8e+06 & 170 & 1.0e+04 & 4.4 & 178 & 40 \\
464 & 22 & 40 & 23 & 57 & 7.0 & 2.7e+05 & 103 & 2.6e+03 & 3.3 & 178 & 54 \\
468 & 54 & 29 & 55 & 45 & 28.4 & 1.0e+07 & 98 & 1.0e+05 & 1.9 & 175 & 92 \\
490 & 82 & 36 & 78 & 49 & 25.7 & 1.2e+07 & 120 & 1.0e+05 & 3.0 & 172 & 57 \\
492 & 52 & 30 & 43 & 43 & 15.6 & 2.5e+06 & 86 & 2.9e+04 & 2.7 & 189 & 69 \\
499 & 49 & 31 & 49 & 45 & 12.9 & 1.9e+06 & 173 & 1.1e+04 & 3.8 & 198 & 52 \\
503 & 41 & 29 & 31 & 38 & 20.7 & 3.2e+06 & 184 & 1.7e+04 & 1.5 & 170 & 112 \\
506 & 14 & 209 & 14 & 372 & 20.4 & 1.5e+06 & 278 & 5.2e+03 & 0.7 & 214 & 298 \\
507 & 14 & 156 & 14 & 259 & 14.8 & 7.3e+05 & 157 & 4.7e+03 & 1.0 & 200 & 210 \\
508 & 17 & 134 & 16 & 225 & 22.2 & 1.9e+06 & 122 & 1.6e+04 & 0.7 & 232 & 313 \\
509 & 25 & 111 & 25 & 190 & 16.7 & 1.7e+06 & 316 & 5.3e+03 & 1.5 & 245 & 162 \\
510 & 7 & 128 & 6 & 175 & 31.4 & 1.5e+06 & 37 & 4.1e+04 & 0.2 & 233 & 1143 \\
516 & 32 & 59 & 35 & 78 & 16.5 & 2.2e+06 & 129 & 1.7e+04 & 2.1 & 239 & 114 \\
520 & 29 & 67 & 30 & 86 & 12.8 & 1.2e+06 & 242 & 4.8e+03 & 2.3 & 246 & 106 \\
521 & 32 & 54 & 35 & 78 & 8.3 & 5.6e+05 & 75 & 7.4e+03 & 4.2 & 243 & 58 \\
527 & 64 & 46 & 67 & 64 & 25.9 & 1.1e+07 & 231 & 4.5e+04 & 2.5 & 218 & 86 \\
532 & 28 & 45 & 26 & 63 & 14.0 & 1.2e+06 & 45 & 2.6e+04 & 1.9 & 243 & 131 \\
534 & 50 & 45 & 49 & 61 & 22.1 & 5.6e+06 & 305 & 1.8e+04 & 2.2 & 217 & 99 \\
536 & 44 & 35 & 47 & 55 & 14.5 & 2.3e+06 & 289 & 7.9e+03 & 3.2 & 227 & 71 \\
550 & 31 & 32 & 29 & 42 & 24.0 & 3.9e+06 & 68 & 5.7e+04 & 1.2 & 235 & 196 \\
564 & 76 & 34 & 78 & 43 & 19.2 & 6.7e+06 & 95 & 7.0e+04 & 4.0 & 209 & 52 \\
595 & 24 & 123 & 24 & 231 & 22.7 & 2.9e+06 & 414 & 7.1e+03 & 1.1 & 290 & 273 \\
598 & 20 & 65 & 21 & 102 & 11.5 & 6.5e+05 & 389 & 1.7e+03 & 1.8 & 290 & 162 \\
599 & 25 & 80 & 26 & 115 & 17.5 & 1.9e+06 & 387 & 4.8e+03 & 1.5 & 264 & 178 \\
601 & 15 & 63 & 16 & 94 & 21.5 & 1.7e+06 & 35 & 4.9e+04 & 0.7 & 273 & 373 \\
603 & 25 & 56 & 25 & 81 & 12.3 & 8.9e+05 & 112 & 7.9e+03 & 2.0 & 309 & 154 \\
605 & 26 & 81 & 26 & 114 & 4.3 & 1.1e+05 & 245 & 4.5e+02 & 6.0 & 288 & 48 \\
607 & 23 & 50 & 23 & 71 & 16.0 & 1.4e+06 & 164 & 8.4e+03 & 1.4 & 297 & 210 \\
613 & 11 & 60 & 10 & 76 & 8.9 & 2.0e+05 & 1831 & 1.1e+02 & 1.2 & 251 & 210 \\
616 & 52 & 47 & 54 & 64 & 27.4 & 9.4e+06 & 438 & 2.2e+04 & 1.9 & 259 & 134 \\
621 & 25 & 59 & 27 & 80 & 12.0 & 9.4e+05 & 63 & 1.5e+04 & 2.3 & 291 & 128 \\
624 & 50 & 38 & 52 & 62 & 22.7 & 6.3e+06 & 550 & 1.2e+04 & 2.3 & 302 & 133 \\
663 & 39 & 30 & 42 & 43 & 18.1 & 3.2e+06 & 131 & 2.4e+04 & 2.3 & 267 & 117 \\
676 & 12 & 147 & 13 & 281 & 18.8 & 1.1e+06 & 113 & 9.4e+03 & 0.7 & 370 & 546 \\
677 & 11 & 159 & 11 & 265 & 18.7 & 9.3e+05 & 65 & 1.4e+04 & 0.6 & 371 & 618 \\
678 & 32 & 113 & 32 & 185 & 15.5 & 1.8e+06 & 495 & 3.6e+03 & 2.0 & 327 & 162 \\
679 & 17 & 80 & 16 & 144 & 19.4 & 1.5e+06 & 84 & 1.7e+04 & 0.8 & 339 & 402 \\
680 & 15 & 64 & 17 & 91 & 7.3 & 2.2e+05 & 34 & 6.3e+03 & 2.3 & 364 & 155 \\
683 & 19 & 61 & 19 & 87 & 18.7 & 1.6e+06 & 129 & 1.2e+04 & 1.0 & 358 & 355 \\
684 & 14 & 66 & 15 & 113 & 11.2 & 4.5e+05 & 141 & 3.2e+03 & 1.3 & 364 & 272 \\
685 & 36 & 73 & 36 & 99 & 17.7 & 2.6e+06 & 404 & 6.6e+03 & 2.0 & 355 & 177 \\
695 & 19 & 47 & 18 & 76 & 34.7 & 5.3e+06 & 86 & 6.1e+04 & 0.5 & 354 & 666 \\
702 & 36 & 64 & 36 & 92 & 13.9 & 1.6e+06 & 68 & 2.4e+04 & 2.6 & 347 & 135 \\
704 & 19 & 67 & 18 & 94 & 8.0 & 2.7e+05 & 80 & 3.4e+03 & 2.2 & 326 & 147 \\
706 & 47 & 52 & 50 & 65 & 17.8 & 3.7e+06 & 236 & 1.6e+04 & 2.7 & 336 & 122 \\
707 & 43 & 53 & 39 & 67 & 24.0 & 5.3e+06 & 89 & 5.9e+04 & 1.6 & 322 & 199 \\
709 & 28 & 36 & 28 & 52 & 15.6 & 1.6e+06 & 93 & 1.7e+04 & 1.8 & 371 & 209 \\
719 & 19 & 36 & 20 & 50 & 11.9 & 6.8e+05 & 61 & 1.1e+04 & 1.7 & 361 & 213 \\
724 & 28 & 40 & 29 & 53 & 10.2 & 7.1e+05 & 170 & 4.1e+03 & 2.8 & 354 & 125 \\
727 & 41 & 36 & 42 & 58 & 39.0 & 1.5e+07 & 186 & 8.1e+04 & 1.1 & 392 & 366 \\
730 & 45 & 39 & 43 & 54 & 20.2 & 4.1e+06 & 281 & 1.5e+04 & 2.1 & 335 & 160 \\
734 & 45 & 31 & 52 & 42 & 25.4 & 7.9e+06 & 62 & 1.3e+05 & 2.0 & 371 & 185 \\
741 & 37 & 28 & 28 & 42 & 15.9 & 1.7e+06 & 112 & 1.5e+04 & 1.7 & 344 & 198 \\
778 & 39 & 36 & 28 & 48 & 10.2 & 6.9e+05 & 563 & 1.2e+03 & 2.7 & 388 & 141 \\
786 & 18 & 108 & 18 & 169 & 25.4 & 2.8e+06 & 152 & 1.8e+04 & 0.7 & 461 & 647 \\
787 & 13 & 90 & 13 & 127 & 23.8 & 1.7e+06 & 266 & 6.5e+03 & 0.5 & 440 & 817 \\
791 & 22 & 83 & 23 & 127 & 19.2 & 2.0e+06 & 101 & 1.9e+04 & 1.2 & 479 & 408 \\
792 & 11 & 65 & 11 & 113 & 25.4 & 1.8e+06 & 129 & 1.4e+04 & 0.5 & 414 & 898 \\
793 & 33 & 80 & 33 & 121 & 11.1 & 9.6e+05 & 112 & 8.6e+03 & 2.9 & 424 & 144 \\
794 & 14 & 94 & 13 & 133 & 21.7 & 1.5e+06 & 46 & 3.2e+04 & 0.6 & 423 & 682 \\
796 & 22 & 83 & 22 & 112 & 14.0 & 1.0e+06 & 115 & 8.7e+03 & 1.5 & 454 & 295 \\
798 & 11 & 111 & 12 & 149 & 28.7 & 2.3e+06 & 31 & 7.4e+04 & 0.4 & 415 & 993 \\
805 & 15 & 67 & 15 & 100 & 17.6 & 1.1e+06 & 37 & 3.1e+04 & 0.9 & 410 & 466 \\
809 & 29 & 50 & 30 & 81 & 16.3 & 1.9e+06 & 115 & 1.7e+04 & 1.8 & 425 & 230 \\
810 & 47 & 46 & 45 & 65 & 14.1 & 2.1e+06 & 178 & 1.2e+04 & 3.2 & 441 & 138 \\
818 & 26 & 39 & 19 & 55 & 11.4 & 5.9e+05 & 53 & 1.1e+04 & 1.7 & 438 & 264 \\
819 & 40 & 48 & 41 & 60 & 15.1 & 2.2e+06 & 133 & 1.6e+04 & 2.7 & 426 & 159 \\
822 & 53 & 42 & 53 & 58 & 14.0 & 2.5e+06 & 426 & 5.8e+03 & 3.7 & 404 & 107 \\
823 & 13 & 47 & 13 & 72 & 12.1 & 4.5e+05 & 254 & 1.8e+03 & 1.1 & 417 & 387 \\
825 & 48 & 41 & 47 & 64 & 30.8 & 1.1e+07 & 117 & 9.0e+04 & 1.5 & 426 & 281 \\
831 & 55 & 32 & 58 & 44 & 30.0 & 1.2e+07 & 194 & 6.4e+04 & 1.9 & 418 & 217 \\
847 & 29 & 32 & 30 & 46 & 6.1 & 2.7e+05 & 289 & 9.4e+02 & 4.9 & 449 & 91 \\
848 & 22 & 27 & 16 & 43 & 8.9 & 3.1e+05 & 958 & 3.2e+02 & 1.8 & 428 & 237 \\
857 & 50 & 37 & 53 & 55 & 8.3 & 8.6e+05 & 408 & 2.1e+03 & 6.3 & 474 & 75 \\
874 & 45 & 25 & 44 & 41 & 35.9 & 1.3e+07 & 132 & 1.0e+05 & 1.2 & 441 & 363 \\
904 & 55 & 30 & 51 & 40 & 12.2 & 1.8e+06 & 339 & 5.3e+03 & 4.2 & 494 & 118 \\
905 & 20 & 120 & 20 & 191 & 21.8 & 2.3e+06 & 121 & 1.9e+04 & 0.9 & 538 & 575 \\
906 & 25 & 142 & 25 & 225 & 19.6 & 2.3e+06 & 340 & 6.8e+03 & 1.3 & 503 & 391 \\
908 & 44 & 89 & 45 & 129 & 16.3 & 2.8e+06 & 329 & 8.4e+03 & 2.7 & 611 & 226 \\
912 & 27 & 50 & 27 & 74 & 14.1 & 1.3e+06 & 109 & 1.2e+04 & 1.9 & 505 & 263 \\
914 & 44 & 39 & 39 & 54 & 25.2 & 5.8e+06 & 52 & 1.1e+05 & 1.5 & 618 & 405 \\
920 & 16 & 37 & 17 & 58 & 19.5 & 1.5e+06 & 66 & 2.3e+04 & 0.9 & 501 & 587 \\
921 & 22 & 55 & 22 & 71 & 15.2 & 1.2e+06 & 28 & 4.4e+04 & 1.5 & 537 & 363 \\
930 & 46 & 38 & 46 & 47 & 19.4 & 4.1e+06 & 176 & 2.3e+04 & 2.3 & 616 & 263 \\
936 & 53 & 54 & 56 & 70 & 11.6 & 1.8e+06 & 111 & 1.6e+04 & 4.7 & 543 & 114 \\
944 & 55 & 44 & 56 & 61 & 13.5 & 2.4e+06 & 517 & 4.6e+03 & 4.1 & 557 & 135 \\
953 & 73 & 42 & 82 & 57 & 35.6 & 2.4e+07 & 78 & 3.1e+05 & 2.3 & 618 & 272 \\
959 & 94 & 26 & 103 & 39 & 17.1 & 7.1e+06 & 109 & 6.5e+04 & 5.9 & 625 & 106 \\
965 & 47 & 32 & 50 & 44 & 13.6 & 2.2e+06 & 56 & 3.9e+04 & 3.6 & 565 & 155 \\
998 & 53 & 29 & 39 & 45 & 3.0 & 8.4e+04 & 52 & 1.6e+03 & 12.6 & 511 & 40 \\
1005 & 6 & 221 & 6 & 358 & 13.9 & 2.9e+05 & 45 & 6.2e+03 & 0.4 & 631 & 1421 \\
1007 & 20 & 87 & 20 & 151 & 28.6 & 3.9e+06 & 186 & 2.1e+04 & 0.7 & 775 & 1092 \\
1008 & 25 & 105 & 24 & 146 & 15.3 & 1.4e+06 & 186 & 7.3e+03 & 1.6 & 732 & 461 \\
1009 & 18 & 61 & 18 & 96 & 15.5 & 1.0e+06 & 29 & 3.5e+04 & 1.2 & 698 & 600 \\
1012 & 24 & 42 & 22 & 66 & 5.5 & 1.6e+05 & 112 & 1.4e+03 & 4.0 & 660 & 163 \\
1014 & 45 & 43 & 42 & 60 & 13.3 & 1.8e+06 & 56 & 3.1e+04 & 3.2 & 756 & 239 \\
1015 & 39 & 40 & 41 & 54 & 24.7 & 5.9e+06 & 100 & 5.9e+04 & 1.6 & 784 & 481 \\
1024 & 27 & 45 & 28 & 58 & 13.2 & 1.2e+06 & 100 & 1.2e+04 & 2.1 & 641 & 300 \\
1028 & 20 & 52 & 21 & 69 & 14.0 & 9.6e+05 & 229 & 4.2e+03 & 1.5 & 641 & 434 \\
1059 & 63 & 28 & 65 & 36 & 14.9 & 3.4e+06 & 42 & 8.0e+04 & 4.3 & 783 & 182 \\
1089 & 16 & 134 & 16 & 207 & 29.5 & 3.4e+06 & 113 & 3.0e+04 & 0.6 & 862 & 1556 \\
1090 & 13 & 158 & 12 & 253 & 15.2 & 6.7e+05 & 23 & 2.8e+04 & 0.8 & 960 & 1208 \\
1098 & 41 & 55 & 40 & 76 & 12.8 & 1.5e+06 & 112 & 1.4e+04 & 3.1 & 947 & 302 \\
1102 & 43 & 37 & 42 & 47 & 15.2 & 2.3e+06 & 45 & 5.1e+04 & 2.7 & 794 & 288 \\
1104 & 32 & 47 & 33 & 90 & 9.0 & 6.4e+05 & 25 & 2.5e+04 & 3.7 & 805 & 220 \\
1107 & 27 & 56 & 29 & 70 & 16.2 & 1.8e+06 & 63 & 2.8e+04 & 1.8 & 963 & 548 \\
1110 & 26 & 47 & 20 & 73 & 15.7 & 1.2e+06 & 34 & 3.5e+04 & 1.3 & 903 & 698 \\
1129 & 26 & 37 & 25 & 51 & 9.2 & 5.1e+05 & 54 & 9.4e+03 & 2.8 & 815 & 296 \\
1181 & 18 & 78 & 17 & 127 & 18.0 & 1.3e+06 & 25 & 5.3e+04 & 1.0 & 1237 & 1279 \\
1183 & 12 & 93 & 12 & 147 & 10.1 & 2.9e+05 & 34 & 8.4e+03 & 1.2 & 1170 & 982 \\
1186 & 30 & 47 & 30 & 66 & 15.5 & 1.7e+06 & 30 & 5.6e+04 & 1.9 & 1135 & 591 \\
1194 & 21 & 60 & 24 & 77 & 13.8 & 1.1e+06 & 31 & 3.4e+04 & 1.7 & 1094 & 626 \\
1198 & 37 & 50 & 35 & 66 & 11.9 & 1.2e+06 & 64 & 1.8e+04 & 3.0 & 1059 & 358 \\
1200 & 32 & 44 & 38 & 69 & 34.5 & 1.1e+07 & 42 & 2.5e+05 & 1.1 & 1159 & 1067 \\
1204 & 31 & 38 & 34 & 52 & 33.3 & 8.8e+06 & 30 & 2.9e+05 & 1.0 & 1023 & 1022 \\
1208 & 43 & 42 & 44 & 58 & 18.7 & 3.6e+06 & 64 & 5.6e+04 & 2.3 & 1082 & 463 \\
1214 & 33 & 40 & 32 & 67 & 17.7 & 2.4e+06 & 369 & 6.4e+03 & 1.8 & 1247 & 685 \\
1216 & 44 & 27 & 58 & 33 & 38.8 & 2.0e+07 & 27 & 7.4e+05 & 1.5 & 1000 & 678 \\
1217 & 48 & 38 & 49 & 49 & 38.9 & 1.7e+07 & 24 & 7.1e+05 & 1.2 & 1040 & 844 \\
1232 & 72 & 37 & 73 & 47 & 12.3 & 2.6e+06 & 229 & 1.1e+04 & 5.8 & 1247 & 214 \\
1243 & 82 & 28 & 90 & 40 & 19.4 & 7.9e+06 & 37 & 2.1e+05 & 4.5 & 1079 & 237 \\
1269 & 50 & 20 & 52 & 27 & 18.7 & 4.3e+06 & 25 & 1.7e+05 & 2.8 & 1139 & 413 \\
1275 & 29 & 125 & 28 & 194 & 16.2 & 1.8e+06 & 126 & 1.4e+04 & 1.7 & 1502 & 859 \\
1278 & 21 & 66 & 23 & 87 & 15.2 & 1.3e+06 & 35 & 3.6e+04 & 1.5 & 1311 & 848 \\
1283 & 41 & 64 & 47 & 72 & 8.2 & 7.3e+05 & 59 & 1.2e+04 & 5.7 & 1443 & 254 \\
1285 & 48 & 36 & 49 & 50 & 15.7 & 2.8e+06 & 165 & 1.7e+04 & 3.1 & 1518 & 496 \\
1286 & 36 & 51 & 37 & 63 & 17.7 & 2.7e+06 & 81 & 3.4e+04 & 2.1 & 1439 & 692 \\
1295 & 63 & 32 & 59 & 44 & 28.2 & 1.1e+07 & 181 & 6.1e+04 & 2.1 & 1544 & 743 \\
1299 & 52 & 30 & 49 & 47 & 14.6 & 2.5e+06 & 77 & 3.2e+04 & 3.3 & 1301 & 391 \\
1300 & 55 & 36 & 58 & 51 & 15.9 & 3.4e+06 & 137 & 2.5e+04 & 3.6 & 1284 & 360 \\
1312 & 80 & 28 & 76 & 42 & 41.0 & 3.0e+07 & 26 & 1.1e+06 & 1.8 & 1309 & 718 \\
1313 & 81 & 42 & 85 & 53 & 12.8 & 3.3e+06 & 43 & 7.5e+04 & 6.5 & 1296 & 198 \\
1315 & 56 & 41 & 54 & 48 & 19.6 & 4.9e+06 & 79 & 6.2e+04 & 2.7 & 1269 & 468 \\
1327 & 65 & 35 & 64 & 42 & 25.3 & 9.6e+06 & 39 & 2.4e+05 & 2.5 & 1488 & 593 \\
1352 & 17 & 56 & 18 & 82 & 32.9 & 4.7e+06 & 26 & 1.8e+05 & 0.5 & 1682 & 3074 \\
1355 & 18 & 54 & 21 & 87 & 21.4 & 2.2e+06 & 32 & 6.8e+04 & 1.0 & 1584 & 1653 \\
1360 & 38 & 47 & 38 & 55 & 17.2 & 2.7e+06 & 35 & 7.6e+04 & 2.2 & 1682 & 764 \\
1362 & 31 & 50 & 33 & 92 & 23.5 & 4.3e+06 & 76 & 5.6e+04 & 1.4 & 1769 & 1286 \\
1377 & 38 & 47 & 38 & 56 & 23.6 & 4.9e+06 & 57 & 8.5e+04 & 1.6 & 1788 & 1130 \\
1378 & 49 & 37 & 49 & 57 & 16.9 & 3.3e+06 & 118 & 2.8e+04 & 2.9 & 1730 & 601 \\
1379 & 22 & 15 & 53 & 18 & 20.0 & 4.9e+06 & 53 & 9.3e+04 & 2.6 & 1603 & 614 \\
1385 & 53 & 34 & 53 & 40 & 13.3 & 2.2e+06 & 100 & 2.2e+04 & 3.9 & 1690 & 433 \\
1394 & 25 & 36 & 25 & 48 & 10.8 & 6.9e+05 & 114 & 6.0e+03 & 2.3 & 1834 & 790 \\
1397 & 51 & 45 & 51 & 59 & 17.5 & 3.7e+06 & 76 & 4.8e+04 & 2.9 & 1651 & 571 \\
1404 & 71 & 37 & 72 & 49 & 22.1 & 8.3e+06 & 361 & 2.3e+04 & 3.2 & 1631 & 506 \\
1415 & 59 & 29 & 61 & 42 & 18.7 & 5.0e+06 & 161 & 3.1e+04 & 3.2 & 1716 & 534 \\
1416 & 63 & 30 & 69 & 36 & 23.3 & 8.8e+06 & 124 & 7.1e+04 & 2.9 & 1876 & 638 \\
1439 & 83 & 31 & 80 & 41 & 13.9 & 3.6e+06 & 82 & 4.3e+04 & 5.6 & 1586 & 280 \\
1455 & 84 & 41 & 78 & 59 & 17.1 & 5.4e+06 & 63 & 8.5e+04 & 4.5 & 1832 & 406 \\
1459 & 49 & 24 & 49 & 31 & 17.8 & 3.7e+06 & 35 & 1.0e+05 & 2.7 & 1839 & 671 \\
1476 & 36 & 88 & 37 & 105 & 29.4 & 7.4e+06 & 31 & 2.4e+05 & 1.2 & 2404 & 1942 \\
1479 & 27 & 32 & 31 & 40 & 18.6 & 2.5e+06 & 26 & 9.5e+04 & 1.7 & 2089 & 1254 \\
1521 & 53 & 36 & 46 & 45 & 12.1 & 1.6e+06 & 47 & 3.3e+04 & 3.8 & 2124 & 559 \\
1528 & 14 & 73 & 14 & 87 & 13.3 & 5.9e+05 & 19 & 3.0e+04 & 1.0 & 2611 & 2489 \\
1532 & 71 & 42 & 72 & 58 & 35.4 & 2.1e+07 & 218 & 9.6e+04 & 2.0 & 2597 & 1301 \\
1535 & 47 & 44 & 46 & 53 & 30.9 & 1.0e+07 & 38 & 2.7e+05 & 1.5 & 2580 & 1750 \\
1540 & 50 & 57 & 48 & 75 & 19.3 & 4.2e+06 & 32 & 1.3e+05 & 2.5 & 2828 & 1153 \\
1542 & 53 & 43 & 46 & 55 & 17.6 & 3.3e+06 & 32 & 1.0e+05 & 2.6 & 2530 & 976 \\
1547 & 59 & 28 & 63 & 34 & 32.2 & 1.5e+07 & 33 & 4.6e+05 & 1.9 & 2967 & 1529 \\
1571 & 13 & 63 & 14 & 97 & 31.4 & 3.3e+06 & 24 & 1.4e+05 & 0.5 & 3570 & 7885 \\
1576 & 45 & 37 & 48 & 46 & 19.6 & 4.3e+06 & 51 & 8.3e+04 & 2.4 & 3213 & 1332 \\
1579 & 60 & 42 & 64 & 54 & 14.4 & 3.1e+06 & 44 & 7.0e+04 & 4.3 & 3559 & 821 \\
1580 & 52 & 38 & 49 & 44 & 27.4 & 8.6e+06 & 87 & 9.9e+04 & 1.8 & 3228 & 1823 \\
1583 & 48 & 41 & 47 & 47 & 15.2 & 2.5e+06 & 25 & 9.9e+04 & 3.0 & 3534 & 1168 \\
1585 & 57 & 32 & 59 & 47 & 20.2 & 5.6e+06 & 48 & 1.2e+05 & 2.9 & 3946 & 1366 \\
1595 & 52 & 31 & 54 & 41 & 21.5 & 5.8e+06 & 31 & 1.9e+05 & 2.5 & 3176 & 1289 \\
1611 & 63 & 60 & 55 & 77 & 21.0 & 5.7e+06 & 38 & 1.5e+05 & 2.6 & 4664 & 1790 \\
1619 & 55 & 44 & 49 & 54 & 21.1 & 5.1e+06 & 34 & 1.5e+05 & 2.3 & 4108 & 1798 \\
1630 & 66 & 58 & 70 & 89 & 33.5 & 1.8e+07 & 30 & 6.1e+05 & 2.1 & 5291 & 2567 \\
1634 & 58 & 46 & 57 & 66 & 11.6 & 1.8e+06 & 28 & 6.5e+04 & 4.9 & 7138 & 1460 \\
1640 & 86 & 31 & 91 & 40 & 20.2 & 8.6e+06 & 25 & 3.4e+05 & 4.4 & 8969 & 2032 \\
\enddata
\tablecomments{Summary of the spatial and dynamical properties of the strings. (1) Name of the string, as given in \citet{Kounkel_2019}. (2) Median \textit{Gaia} DR2 distance offset (in parsecs) between stars in the string and the closest point on the string spine. (3) Median signal to noise of the parallax measurements of the string members in \textit{Gaia} DR2. (4) Median \textit{Gaia} DR3 distance offset (in parsecs) between stars in the string and the closest point on the string spine. (5) Median signal to noise of the parallax measurements of the string members in \textit{Gaia} DR3. (6) Error-weighted radial velocity dispersion of the subset of stars in the string with \textit{Gaia} DR3 radial velocity data, after removing flagged binaries. (7) Predicted virial mass of the string defined using the radial velocity dispersion. (8) The estimated observed mass of the string, computed by counting the number of stellar members and assuming a mean stellar mass of 0.61 $M_\odot$ per member. (9) The ratio of the string's virial mass to its observed mass. (10) The predicted dispersal time of the string given its unbound state, defined as the crossing time (ratio of Column 4 over Column 6). (11) Reported age of the string in \citet{Kounkel_2019} (see their Table 2). (12) Ratio of the reported age of the string over the dispersal time. }
\end{deluxetable*}